\documentclass[12pt,preprint]{aastex}

\shortauthors{Junkkarinen et al.}
\shorttitle{Dust and Diffuse Interstellar Bands}

\begin{document}

\title{Dust and Diffuse Interstellar Bands in the $z_a$ = 0.524
Absorption System toward AO 0235+164\altaffilmark{1,2}}
\author{V. T. Junkkarinen, Ross D. Cohen, E. A. Beaver,
E. M. Burbidge, R. W. Lyons}
\affil{Center for Astrophysics and Space Sciences, University of California, San
Diego, La Jolla, CA 92093-0424}
\email{vesa@ucsd.edu, rdcohen@ucsd.edu}
\author{G. Madejski}
\affil{Stanford Linear Accelerator Center, 2575 Sand Hill Road,
Menlo Park, CA 94025}

\altaffiltext{1}{Based on observations with the NASA/ESA Hubble Space
Telescope, obtained at the Space Telescope Science Institute,
which is operated by the Association of Universities for Research in Astronomy,
Inc. under NASA contract No. NAS5--26555.}
\altaffiltext{2}{Some of the data presented herein were obtained at the
W. M. Keck Observatory which is operated as a scientific partnership
among the California Institute of Technology, the University of California
and the National Aeronautics and Space Administration.
The Observatory was made possible by the generous financial support of
the W. M. Keck Foundation.}

\begin{abstract}
We present new HST STIS NUV--MAMA and STIS CCD observations of the 
BL Lac object AO 0235+164 and the intervening damped Ly $\alpha$ (DLA) 
line at $z_a$ = 0.524.  The line profile gives $N(H I)$ = 5 $\pm$ 1
$\times$ 10$^{21}$ cm$^{-2}$ and, combined with the H I 21 cm absorption
data leads to a spin temperature of $T_s$ = 220 K $\pm$ 60 K.  Those 
spectra also show a strong, broad feature at the expected position of 
the 2175~\AA \ graphitic dust feature at $z_a$ = 0.524.  Assuming 
a Galactic type dust extinction curve at $z_a$ = 0.524 gives a dust--to--gas 
ratio of 0.19 Galactic, but the fit, assuming the underlying, 
un--reddened spectrum is a single power--law, is poor in the far--UV.  A dust--to--gas 
ratio of 0.19 Galactic is similar to the LMC, but the AO 0235+164 
spectrum does not fit the LMC extinction curve, or the SMC extinction 
curve (which has practically no 2175~\AA \ feature).  A possible 
interpretation includes dust similar to Galactic, but with less of the 
small particles that produce the far--UV extinction.  The metallicity of 
the $z_a$ = 0.524 absorber, estimated from the observed $N(H I)$ 
and excess X-ray absorption (beyond Galactic) derived from contemporaneous 
and archival ASCA and ROSAT X-ray data, is $Z$ = 0.72 $\pm$ 0.28 $Z_{\sun}$, 
implying in turn the dust--to--metals ratio of 0.27 Galactic.  
If the dust mass density is the same in the $z_a$ = 0.524 DLA system 
as in our Galaxy, only 14\% ($\pm$ 6\%) of the metals (by mass) 
are in dust compared with
51\%, 36\%, and 46\% for the Galaxy, LMC, and SMC respectively.
Such a dusty $z_a$ = 0.524 AO 0235+164 absorption system is a good
example of the kind of DLA system that will be missed
due to selection effects, which in turn can bias the measurement
of the co-moving density of interstellar gas (in units of the closure 
density), $\Omega_g$, as a function of $z$.

\end{abstract}


\section{Introduction}

Dust in damped Ly $\alpha$ (DLA) systems has important implications
for studies of the chemical evolution of the universe \citep{pei1995}.
Dust absorption is correlated with metallicity so the more
metal rich systems will be underrepresented in
DLA surveys using magnitude limited samples of background targets.
Evidence for dust in DLA systems comes from the statistics of
the spectral slopes of the background quasars.  Quasars with
DLA systems have, on average, redder 
spectra \citep{fallpei1989,fallpeimcmahon1989}.
The background control quasars have a range of spectral slopes, so the
confirmation of reddening in any individual DLA system is usually not possible
by this method.

Evidence for dust in DLA systems also comes from studies of the column
density ratios of certain ions, in particular Cr II/Zn II
\citep{pettini1994,pettini1997a}.  Cr is more readily depleted by
grain formation than Zn which normally undergoes very little
dust depletion.
\citet{pettini1997a,pettini1999} find low metallicities 
based on Zn, with [$\langle$Zn/H$\rangle$]\footnote{Where
[A/H] is the base 10 logarithm of the abundance ratio of 
A/H divided by the solar A/H ratio.}
 = $-$1.03 for $z$ $<$ 1.5,
a large scatter, and little or no evolution of metallicity with redshift.
The dust content in 
these low metallicity systems is low mainly because the systems
are low metallicity but also because the available metals are not efficiently
converted to dust.   Based on Cr II/Zn II and other line ratios,
the dust--to--metals ratio 
is estimated to be $\sim$ 0.5 times that of the local Galactic ISM 
\citep{pettini1997a}.  The combination of low metallicity and the
0.5 fractional dust--to--metals ratio relative to Galactic
leads to the result that dust should mainly show up 
in DLA systems in the selective depletion of certain elements,
like Cr, while 
the typical extinction due to dust is small, $A_{1500}$ $\sim$ 0.1 
\citep{pettini1997a}.  However, there is still some uncertainty in the 
interpretation of the observed element abundances in DLA systems because 
the overall pattern may not match the combined effect of early $\alpha$ 
process (SNe II) enhancement combined with dust depletion 
\citetext{e.g. \citealp{lu1996,prochaska1999}}.  There may be a problem 
with the use of Zn as a metallicity indicator because super--solar
values of [Zn/Fe] might result from early Zn production by SNe II 
nucleosynthesis \citep{prochaska2000}.  If Zn is not a good metallicity 
indicator, then [Cr/Zn] is an unreliable indicator of dust
when Cr is apparently depleted by only a small amount relative to Zn.

A more direct detection of dust in individual DLA systems is
possible by the observation of the 2175~\AA \ dust feature thought 
to be produced by graphitic dust grains 
\nocite{draine1993,will1999}(Draine 1993; Will and Aannestad 1999).
Somewhat less direct evidence for dust is the presence of diffuse
interstellar bands (DIBs) which are known to correlate with $E(B - V)$
and other dust indicators in our Galaxy \citetext{e.g. \citealp{herbig1995}}.
DIBs have been observed in only a few extragalactic objects:
the LMC (Vladilo et al. 1987) and NGC 5128 (Cen A) (D'Odorico et al. 1989)
and toward some dusty starburst galaxies \citep{heckmanlehnert2000}.

A search for the 2175~\AA \ feature in a sample of quasars with DLA
systems by \citet{pei1991} yielded no detections.  Pei et al. suggest 
that Galactic type dust is ruled out in these systems and that the 
dust absorption may have an extinction curve similar to that associated 
with either the Small Magellanic Cloud or the Large Magellanic Cloud.  
The SMC extinction has essentially no 2175~\AA \ feature while the LMC 
extinction has a weaker 2175~\AA \ extinction feature relative to 
Galactic extinction \citetext{e.g. \citealp{pei1992}}.

A broad 2175~\AA \ feature has been detected by \citet{malhotra1997} in
a composite spectrum created by aligning 96 quasar Mg II absorption
systems.
The nature of the composite spectrum and the relatively shallow feature
found by \citet{malhotra1997} make any dust--to--gas ratio and extinction
curve determination very uncertain.
A Galactic type 2175~\AA \ feature has recently been discovered in a quasar
absorption system at $z_a = 0.83$ by \citet{motta2002}.  This system
was found in the analysis of the spectra of the gravitationally lensed
quasar SBS 0909$+$532. The 2175~\AA \ feature is apparent in the ratio
of the spectra from the two lines--of--sight.  Such lines--of--sight are 
very valuable for determining extinction curves because the extinction 
can be determined independently of the underlying spectral shape.
Because the absorption is probably due to the lensing galaxy 
\citep{motta2002}, the selection properties of these systems differ 
from the absorption survey selected high column density and DLA absorbers.

AO 0235+164 is a well 
studied BL Lac object that shows high amplitude optical variability 
\citep{rieke1976}, rapid X--ray variability \citep{madejski1996},
strong GeV $\gamma$--ray emission \citep{hunter1993},
and possible VLBA radio components that move with apparent 
superluminal expansion velocities \citep{marscher1999}.
AO 0235+164 has characteristics consistent with a fairly extreme
example of a blazar, which are thought to be AGN with a jet aligned
closely to our line--of--sight \citetext{e.g. \citealp{urry1999}}.
More specifically, given that AO 0235+164 was selected by radio emission
and has a relatively low ratio of X-ray flux to 100 Mev $\gamma$-ray
flux \citep{madejski1996}, AO 0235+164 is probably best described as 
a ``red'' or, equivalently, a low--frequency peaked BL Lac object
\citetext{e.g. \citealp{urry1999}}.  In its low state AO 0235+164 
shows weak emission lines at $z_e$ = 0.94 \citep{cohen1987}.  
Sometimes, when the continuum emission is very weak, the emission line 
equivalent widths are large enough that AO 0235+164 could be classified 
as an optically violent variable (OVV) quasar \citep{nilsson1996}.

During a particularly bright flare in 1975, when AO 0235+164 reached 
V = 14.3 \citep{rieke1976}, optical observations showed absorption 
systems at $z_a$ = 0.524 and $z_a$ = 0.852 \citep{rieke1976, burbidge1976}.  
The $z_a$ = 0.524 absorption system has a high H I column density;  
strong multicomponent H I 21 cm absorption was detected 
against the background radio source \citep{roberts1976}.  
The 21 cm H I absorption is time variable \citep{wolfe1982}.  
The radio observations combined with a measurement of the H I 
column density from the DLA line at 
$\lambda_{obs}$ = (1 + $z_a$) 1216~\AA \ = 1853~\AA \ 
can be used to estimate the spin temperature $T_s$ of the gas producing 
the absorption.  The measurement of the DLA line has been the goal 
of several investigations.  The DLA line at $z_a$ = 0.524 was observed 
with IUE yielding $N(H I)$ = 3 $\times$ 10$^{21}$ cm$^{-2}$ with factor 
of 2 errors from a very noisy spectrum \citep{snijders1982}.
The HST FOS GTO group at UCSD attempted to obtain a higher S/N UV 
spectrum of AO 0235+164 in July 1995.  At the time of the attempted 
observations, AO 0235+164 had varied to a historically faint
apparent brightness around V = 20.5 \citep{burbidge1996}.  The field of 
AO 0235+164 has a Seyfert type emission line galaxy, called object A, 
with $z_e$ = 0.524 about 2 arcsec south of the BL Lac object 
\citep{smith1977}.  The FOS target acquisition aperture
included both AO 0235+164 and object A, and object A was brighter.
The end result of the July, 1995 FOS observation was a spectrum 
of object A which shows not only strong AGN type emission
lines but also broad associated absorption \citep{burbidge1996}.

In this paper we report results from a GTO HST STIS program (GO--7294, PI
R.D. Cohen) and Asca observation to obtain UV, X-ray 
and optical spectra of AO 0235+164.
These spectra, obtained in February, 1998 when AO 0235+164 was 
relatively bright, V $\sim$ 16.4, not only provide a measurement 
of $N(H I)$ from the DLA line at $z_a$ = 0.524, but also reveal 
a strong 2175~\AA \ graphitic dust feature at $z_a$ = 0.524.  
This is the first direct detection of the 2175~\AA \ feature
in a quasar DLA system.  We also report on a Keck LRIS spectrum 
of AO 0235+164 that shows a weak 4428~\AA \ diffuse interstellar 
band (DIB) feature at $z_a$ = 0.524, the first detection of a 
DIB feature in a quasar DLA system.  

The measured $N(H I)$ from the STIS spectrum and concurrent Asca
and previous Asca and ROSAT
X--ray spectra are used to find the metallicity 
in the $z_a = 0.524$ DLA system toward AO 0235+164.
These results are compared to a recent determination of the metallicity
in this system based on a Chandra X--Ray Observatory ACIS-S spectrum of
AO 0235+164 \citep{turnshek2003}.

This paper is organized as follows.  The observations section, \S2, 
describes the spectra that are analyzed. Section \S2.1 describes
the HST STIS spectra and \S2.2 presents the Keck Observatory
data.  Section \S2.3 describes the
simultaneous Asca data and spectral fits to all the X--ray
data.  The analysis section, \S3, 
is divided into four subsections: \S3.1, the determination of $N(H I)$ 
from the DLA line and the spin temperature, \S3.2, the metallicity,
\S3.3, the measurement of dust properties from the 
2175~\AA \ feature, and \S3.4, the measurement of the DIB feature.  The 
analysis is followed by a discussion section, \S4, which covers the 
implication of these observations regarding the nature of dust and 
selection effects in DLA systems.  This paper concludes with a brief
summary, \S5.

\section{Observations}

\subsection{HST/STIS Data}

AO 0235+164 was observed with HST/STIS on 11 February 1998 (UT) in
the spectroscopic mode.  The observations consisted of STIS 
NUV-MAMA G230L (1570--3180~\AA ), STIS CCD G430L (2900--5700~\AA ),
and STIS CCD G750L (5236--10266~\AA ) spectra, with total integration times
of 12,833 s (5 orbits), 2874 s (1 orbit), and 2160 s (1 orbit) respectively.
All of the data were obtained using the 
0.5 arcsec wide slit.
The data were reduced using STScI STSDAS software. 
The G750L observation was flat fielded using a contemporaneous
flat field exposure to correct fringing.
A comparison of the spectrum before and after the fringing correction
shows that the corrected spectrum
is free from apparent structure due to residual fringing.
Before performing a $\chi^2$ analysis on these three spectra,
the flux-corrected STIS G230L spectrum was
adjusted by multiplication by 0.98 to match the G430L spectrum 
where the spectra overlap.

\subsection{Keck Data}

Observations of AO 0235+164 from the Keck Observatory were obtained
as part of a program to study the galaxies near AO 0235+164 (P.I. E. M. 
Burbidge).  The observations were obtained on 1997 November 7 (UT)
with the Keck II 10 m telescope and the LRIS low resolution spectrograph.
A 300 g/mm 5000~\AA \ blaze grating provided a wavelength coverage of
4000~\AA \ to 9000~\AA \ (with order overlap) and a resolution (with a
1.0 arcsec wide slit) of about 10~\AA \ FWHM and 2.5~\AA \ per pixel sampling.
The spectrum of AO 0235+164 discussed in this paper is from an 1800 s
integration at a slit PA = 0 degrees that was chosen to include
AO, a faint galaxy due north of AO, and object A.  The integration 
begun at HA = 1:05 East and the beginning airmass was 1.02.  The data were 
reduced using VISTA.

Figure 1 shows the HST STIS and Keck LRIS spectra.
The 1.0 arcsec wide slit Keck LRIS observation
of AO 0235+164 was scaled by 1.58, the ratio of fluxes between 
an 8.7 arcsec wide and 1.0 arcsec wide slit observation
of Feige 110, the standard star observed at the beginning of night.
Because slit losses can vary depending on guiding and seeing, the absolute
flux of the 7 November 1997 Keck spectrum of AO 0235+164 has a large
uncertainty, of order $\pm$ 20\% . 
The Keck observations during the short, half night,
program were intended to measure emission lines in faint nearby galaxies,
especially [O II] $\lambda$3727 at $z_e$ = 0.524, and were not particularly
well suited for high signal--to--noise spectra of AO itself.
The Keck spectrum of AO 0235+164 shows structure in the 7000 to 
9000~\AA \ range due to less than perfect fringe correction and 
residuals from atmospheric absorption bands at about 6900~\AA \ 
and 7600~\AA .
A comparison of the STIS and Keck spectra scaled to match each other around
6000~\AA \ shows that the Keck spectrum is somewhat steeper than the HST 
STIS spectrum below 7000~\AA .  The difference is marginally significant 
since it amounts to $\pm$ 5\%, which is about the maximum uncertainty 
that is expected in the relative flux calibration for the Keck LRIS spectrum.
At wavelengths longer than 7000~\AA \ the Keck spectrum is not reliably
flux calibrated due to order overlap.  The highly variable AO 0235+164 
was about a factor of 2 brighter on 11 February 1998 during the
STIS observations than on 7 November 1997, some 4 months earlier. 

\placefigure{fig1}

\subsection{Asca Data and Spectral Fitting}

Asca observation of AO 0235+164 started on February 11, 1998 at 7:58 UT.  
The observation lasted for about 16 hours, and yielded about 15 ks 
of useful data.  The data were reduced in a standard manner, where 
the screening criteria were the same as those used in the reduction 
of the 1994 Asca data for this object by \citet{madejski1996}.  
In our analysis, we used the data reduction tools provided by
HEASARC; the most recent tools available as of January 1, 2003.  
This includes the standard response matrices for GIS detectors, 
and matrices derived from standard calibration files via the 
{\tt sisrmg} tool for the SIS detectors.  The effective area files 
were calculated using the {\tt ascaarf} tool v. 3.10 using the 
standard Asca calibration files.  All reduction of the SIS data 
was performed in the BRIGHT2 mode.  For further analysis, the data 
were binned such that there were at least 20 counts in each new 
energy bin, and the energy range for spectral fits was restricted
to the bandpass of 0.5 - 10 keV for all four instruments.  For all 
four detectors, the background was extracted from the same image as 
the source, avoiding any obvious point sources.  The background-subtracted 
source count rates were respectively $0.109 \pm 0.003$, $0.089 \pm 0.003$, 
$0.065 \pm 0.002$ and $0.084 \pm 0.002$ counts s$^{-1}$ in SIS0, SIS1, 
GIS2, and GIS3 detectors.  We saw no measurable variability of the 
count rate within the span of our observation.  Note that those count 
rates were roughly 2.5 times greater than the count rates during the 
1994 Asca observation.  

The spectral fitting of the data was entirely analogous to that 
reported in \citet{madejski1996}:  data from all four detectors 
were fitted to a single common model, including a power law continuum, 
modified by photoelectric absorption.  In agreement with the previous 
spectra, the inferred column was in excess of the Galactic value.  
We fixed the Galactic column to $7.6 \times 10^{20}$ cm$^{-2}$ 
\citep{elvis1989}, and assumed that the 
excess would be due to the intervening 
system at $z = 0.524$.  For the purpose of comparison with the
previous Asca data sets, we assumed (incorrectly, as we argue below) 
that the entire excess absorption (beyond the Galactic value) 
is at $z = 0.524$ and obeys solar abundances.  
With this, the equivalent hydrogen column density at $z = 0.524$ 
was fitted to be $3.4 \pm 1.1 \times 10^{21}$ cm$^{-2}$, which, within 
errors, was consistent with the values measured via previous Asca and ROSAT 
observations.  The energy power--law index $\alpha$ was $0.72 \pm 0.07$
and the $\chi^{2}$ for the 1998 Asca data was 260 for 263 PHA bins. 
This indicates that the spectrum was harder in 1998 than in 1994 when 
$\alpha$ was $0.93 \pm 0.07$, but this was not surprising, as the indices already 
differed between the previous Asca and ROSAT observations, conducted 
$\sim 5$ months apart.  
As mentioned above, the source was brighter than in the 1994 Asca 
observation:  the 2 - 10 keV model flux for the Feb. 1998 data 
was $4.1 \times 10^{-12}$ erg cm$^{-2}$ s$^{-1}$, vs. 
$1.2 \times 10^{-12}$ erg cm$^{-2}$ s$^{-1}$ in Feb. 1994.
Figure 2 shows the 1998 Asca X-ray spectrum of AO 0235-164.

\placefigure{fig2}
 
Since the absorption values inferred from the ROSAT and the two Asca data 
sets were consistent with a single value, in our subsequent analysis we 
fitted all three data sets simultaneously, assuming constant effect 
of the combined absorption.  We assumed that the
intrinsic spectrum of the BL Lac object is a simple power law, 
but we allowed the normalization and the power law index to vary 
from one data set to another to account for the observed continuum 
variability.  We used the 
ROSAT and 1994 Asca data reduced in a manner given in \citet{madejski1996}.  
In addition, we performed the analysis using the current calibration 
files for Asca and ROSAT, but there was no discernible difference 
between the two analyses.  As above, we fixed the Galactic absorption 
at $7.6 \times 10^{20}$ cm$^{-2}$.  Regarding the modeling of the 
intervening absorption, we followed the 
approach of \citet{madejski1996}, where we assumed that the absorption 
at $z = 0.524$ consists of two components, both parameterized 
via equivalent hydrogen column density.  One component, having the primordial 
composition of elements, consists of pure hydrogen plus 76 \% of the 
solar abundance value of helium (by number of atoms).  The other component 
represents all processed elements with their relative ratios to each
other at the solar value, but it contains no hydrogen, and only 24 \% 
of the solar abundance value of helium.  Such an approach is necessary, 
since the helium in the primordial absorber has an effect on the 
absorption even in the Asca band.  It also allows us 
to immediately estimate the metallicity of the intervening absorber 
since the ratio of the processed column to the column inferred from 
the damped Ly $\alpha$ analysis yields the metallicity of the intervening 
system in solar units.  

\placefigure{fig3}

Given the modest energy resolution of the Asca or ROSAT detectors, no discrete 
spectral features were detected in any data sets.  With this, the 
relative contributions of either absorption component are highly correlated, 
and this is clearly illustrated in Figure 3.  While the best-fit values of 
the primordial and processed columns would be respectively  
$10.5 \times 10^{21}$ cm$^{-2}$ and $2.3 \times 10^{21}$ cm$^{-2}$, 
yielding $\chi^{2}$ of 528.5 for 535 PHA bins, fixing the primordial 
column at $5 \times 10^{21}$ cm$^{-2}$ (based on the DLA fit -- see \S3.1)
increases $\chi^{2}$ only to 
530.0, and implies that the column density of the processed material is
$3.6 \pm 0.8 \times 10^{21}$ cm$^{-2}$.  With this, we immediately find 
the metallicity of the intervening system, $Z$, to be $3.6/5 = 0.72 \pm 0.24 $ 
solar (the quoted uncertainty includes the range of X-ray measurements 
and the uncertainty of the H I column based on the STIS spectrum, added in quadrature).  

\section{Analysis}

\subsection{H I Column Density and Spin Temperature}  

Figure 4 shows the STIS NUV MAMA G230L spectrum of AO 0235+164
in the region around the DLA line at $z_a$ = 0.524 and a $\chi^2$
fit to G230L spectrum from 1690 \AA \ to 2240 \AA .  This fit gives 
$N(H I)$ = 5 $\pm$ 1 $\times$ 10$^{21}$ cm$^{-2}$.
The one $\sigma$ uncertainty includes both the 
formal error and an estimate of some of the systematic
uncertainties.  The one $\sigma$ formal error
in $N(H I)$ from the $\chi^2$ fit and
the associated curvature matrix is ($+0.6,-0.5$)
$\times$ 10$^{21}$ cm$^{-2}$.
The continuum level, continuum slope, and $E(B - V)$ at $z$ = 0.0 
were left as free parameters in this $\chi^2$ fit so the uncertainties 
in these values entered the $N(H I)$ formal uncertainty in the proper 
way from the cross terms in the curvature matrix.  Possible systematic 
errors result from the unknown shape of the local continuum.
The continuum may deviate from a power law, and there are uncertainties
in the shape of the $z_a$ = 0 and $z_a$ = 0.524 dust absorption
extinction curves.  Weak metal lines in the $z_a$ = 0.524 and $z_a$ = 0.852
absorption systems occur in the wings of the DLA profile
along with possible Ly$\alpha$ forest contamination with some
uncertainty in the relative line strengths.
These systematic errors were investigated by repeating the fitting 
process assuming different shapes for the continuum and different 
line strengths for the weak lines.
The relative strengths of all the weak lines 
are not completely arbitrary.
Each line, modeled as a Voigt profile with a single
component, has a strength determined by the column density of
the ion producing the line.
Solar abundances were sometimes used as a guide to determine
ranges for column density ratios between selected ions.
For example, the Si II column density is constrained to be less than
the C II column density in the model for the absorption lines produced
by the $z_a$ = 0.524 DLA system.
Some lines, such as those from Si II, are constrained by multiple 
lines from Si II within this spectral range.
Other absorption lines, including the lines that are Ly $\alpha$ forest lines, are
only constrained by the data.
Repeating the $\chi^2$ fit after
excluding from the model spectrum all the weak lines that are not 
well detected, and treating all of the observed structure on the wings 
of the DLA line as noise, results in a higher $N(H I)$ estimate;
$N(H I)$ = 7 $\times$ 10$^{21}$ cm$^{-2}$.  But this model
spectrum does not fit the core of the observed
DLA line as well as the $N(H I)$ = 5 $\times$ 10$^{21}$ cm$^{-2}$ model.
Trials with different parameters for the weak metal lines and different
values of $R_V$ = $A_V / E(B - V)$ at $z$ = 0 and $z$ = 0.524 lead to a
final estimate of $N(H I)$ = 5 $\pm$ 1 $\times$ 10$^{21}$ cm$^{-2}$.

\placefigure{fig4}

The best fit value for Galactic reddening, $E(B - V)$ at $z$ = 0.0,
from the fit shown in Figure 4, is $E(B - V)$ = 0.13.  The 
Galactic 21 cm H I emission data 
give $N(H I)$ = 7.6 $\times$ 10$^{20}$ cm$^{-2}$ 
in the direction of AO 0235$+$164 \citep{elvis1989}.  
Using the average Galactic $N(H I)$ to $E(B - V)$ 
relationship from \citet{diplas1994}, 
$N(H I)$ = 4.93 $\times$ 10$^{21}$ cm$^{-2}$ $E(B - V)$,
gives $E(B - V)$ = 0.154, very close to the $E(B - V)$ = 0.13
from the $\chi^2$ fit used to find $N(H I)$.
Repeated trials with different continuum assumptions show that
the best fit $E(B - V)$ value at $z$ = 0.0 is not well determined 
from the data because the 2175~\AA \ dust feature at $z$ = 0.0 is 
relatively shallow, the unreddened continuum slope is not known, 
and the statistical errors in the STIS NUV spectrum of AO 0235+164 
at the bluest wavelengths are large.  In $\chi^2$ fits to measure 
the 2175~\AA \ dust feature at $z_a$ = 0.524, described in the next
section, the $E(B - V)$ at $z$ = 0.0 was fixed at $E(B - V)$ = 0.154.
Setting $E(B - V)$ = 0.154 at $z$ = 0 in the above fit gives
a log $N(H I)$ = 21.72 (5.2 $\times$ 10$^{21}$ cm$^{-2}$) at $z_a$ 
= 0.524 which is only 0.02 dex higher than the value above, and within 
the formal $1 \sigma$ errors.

Combining the DLA determined $N(H I)$ value with the H I 21 cm 
absorption observation from \citet{roberts1976} who found
$N(H I)$ = 2.3 $\times$ 10$^{19}$ $T_s$ cm$^{-2}$ gives
$T_s$ = 220$\pm$60 K.  The Roberts et al.~result assumes complete
covering of the background source
and is an integration over the entire observed 21 cm feature
with the same $T_s$ assumed for all the absorption components.
For comparison, $T_s$ has been estimated in a few DLA 
systems \citetext{e.g. \citealp{briggs1987}; \citealp{cohen1994}}
and the range of values is $T_s$ = 600 K to $T_s$ = 1500 K.
There are several uncertainties in deriving $T_s$ from
spectroscopic DLA measurements and radio 21 cm absorption.
The background radio source and optical/UV source do not coincide
for many quasars \citetext{e.g. \citealp{briggs1999}}.
In the case of the $z_a$ = 0.524 DLA system toward AO 0235+164,
the background radio source is very compact.
Thus the $T_s$ estimate for the AO 0235$+$164 $z_a$ = 0.524 DLA system
may be somewhat more robust than the $T_s$ estimates in some other 
DLA systems.  However there is an added uncertainty for AO 0235$+$164
because the $z_a$ = 0.524 H I 21 cm absorption is time variable 
\citep{wolfe1982}.   An inspection of the 14 epochs of H I 21 cm absorption
displayed by \citet{wolfe1982}, obtained over a four year period
from 1977 to 1981, shows that, although there are large
(up to a factor of two) changes in individual components,
the added uncertainty to $T_s$ due to variability is not large, $\le$ 20\%
peak to peak, if the 21 cm absorption stays within the range observed by
\citet{wolfe1982}.  Other systematic errors in $T_s$ are more 
difficult to estimate and the $\pm$60 K uncertainty given above 
reflects only the error in $N(H I)$ combined in quadrature
with $\pm$20\% as a rough estimate for the effects of variability.

\subsection{Metallicity}

As discussed above in \$2.3, the total metallicity in the $z_a$ = 0.524 DLA 
system can be estimated from the observed $N(H I)$ and X-ray absorption.
The main elements contributing to the X-ray absorption in the ROSAT / 
Asca bands at the column inferred here are oxygen, and to a lesser extent, 
carbon and iron.  The X-ray absorption measurement is particularly valuable, 
as the inferred column includes a contribution of all phases 
of the the absorber, not just the gas phase.  
The observed $N(H I)$ = 5 $\times$ 10$^{21}$ cm$^{-2}$ from the DLA fixes
the hydrogen column in the ``primordial'' component.
With a fixed ``primordial'' component, the spectral fit to the X-ray data 
above gives a metallicity $Z$ = 0.72 $\pm$ 0.24 $Z_{\sun}$, where again, 
the range of allowed X-ray absorption and the uncertainty in $N(H I)$ 
are added in quadrature.  There is some added uncertainty due to the 
time variable nature of the absorber given our use of X-ray data which 
are at least in part, non-simultaneous with the STIS data.
Adding in a 20\% additional uncertainty due to variability
(a rough estimate based on the maximum range of 21-cm
variability observed by \citet{wolfe1982} -- see above) increases the
uncertainty to $\pm$ 0.28.  So our final metallicity is
$Z$ = 0.72 $\pm$ 0.28 $Z_{\sun}$.

It is interesting to juxtapose the comparison of the H I absorption 
data and the X-ray column for AO 0235+164 against those measured in 
another DLA system at slightly lower redshift, $z_{abs} = 0.312$, 
located in front of PKS 1127-145 (which is at $z_{em} = 1.187$).  
This system shows an H I column of $5.1 \pm 0.9 \times$ 10$^{21}$ 
cm$^{-2}$, quite similar to the column in AO 0235+164.  Chandra 
observations of this object revealed excess X-ray absorption above 
the expected Galactic value \citep{bechtold2001}.  However, this 
excess is significantly smaller than in AO 0235+164, indicating that 
the metallicity is less than 0.17 $Z_{\sun}$, 
much lower than in AO 0235+164.  This implies that the 
metallicity of the ISM varies significantly from one galaxy 
to another.  We also note that the measurements in both cases are performed in 
a single line--of--sight, covering at most a patch corresponding to a 
square parsec of the intervening system, and the ISM of any given galaxy 
might well show variation of metallicity from one line--of--sight to another.  

AO 0235+164 has also been observed with the Chandra X--ray Observatory
\citep{turnshek2003}
to determine the metallicity in the $z_a$ = 0.524 DLA
system.  \citet{turnshek2003} use the HST STIS spectrum discussed in this
paper to obtain an H I column density and they find $N(H I) = 4.5 \times 10^{21}$ cm$^{-2}$
$\pm 0.4 \times 10^{21}$ cm$^{-2}$.
As in this paper, the observed X--ray spectrum is assumed to be a power law
with absorption
at $z_a = 0$ and $z_a = 0.524$.
\citet{turnshek2003} find a metallicity of
$Z = 0.24 \pm 0.06 Z_{\sun}$
using a solar metallicity model for the Galactic absorption and solar
abundance ratios for the metals in the
absorber at $z_a = 0.524$.
\citet{turnshek2003} also give metallicities
assuming a Galactic
ISM with no metals and with reduced metals from solar and assuming that
the absorber at $z_a = 0.524$ has
enhanced $\alpha$ process metals by a factor of 2.5 relative to solar.
The six metallicity values found by \citet{turnshek2003}
for the AO 0235+164 $z_a = 0.524$ DLA system, depending on the assumptions, range
from $Z = 0.11 \pm 0.03 Z_{\sun}$ to $Z = 0.61 \pm 0.07 Z_{\sun}$.
The \citet{turnshek2003} solar metallicity model at $z_a = 0.524$ and solar metallicities
for the Galactic ISM are similar to the assumptions used in this paper.
We find $Z = 0.72 \pm 0.28 Z_{\sun}$ using these assumptions and the combined ROSAT / Asca data
sets; a metallicity
that is 3 times as large as the \citet{turnshek2003}  value
of $Z = 0.24 \pm 0.06 Z_{\sun}$.
The formal difference between the metallicity found in this
paper and that found by \citet{turnshek2003} is
about 2.0$\sigma$ for the quoted uncertainties before we add the systematic effect
due to possible time variability.
The effect of the slightly different input values for $N(H I)$ is only 10\% and using
the value found in this paper,
$N(H I) = 5 \times 10^{21}$ cm$^{-2}$, would decrease the metallicity
of \citet{turnshek2003} by 10\%.
The Galactic column density toward AO 0235+164 used by \citet{turnshek2003} is
$N(H I) = 8.7 \times 10^{20}$ cm$^{-2}$ (based on 21 cm maps)
compared to the column density used in this
paper of $N(H I) = 7.6 \times 10^{20}$ cm$^{-2}$ (based on a pointed 21 cm observation
toward AO 0235+164 - \citet{elvis1989}).
Given the nature of the 21 cm observations to 
determine the Galactic $N(H I)$ values and the quoted errors
in the 21 cm H I column densities,
either value of Galactic column density is possible.
The higher value Galactic column used by \citet{turnshek2003}
would tend to give a somewhat lower total absorption at $z_a = 0.524$.
Most of the difference in the metallicities between
the Asca and ROSAT X--ray data analyzed in this paper and
the Chandra X--ray data analyzed by \citet{turnshek2003} 
is probably due to different realizations of the noise in the
X--ray data sets or due to time variability or some combination of the two effects.
Additional differences may result if a single power law is
not an accurate representation of the unabsorbed X--ray spectrum of AO 0235+164.
Future X--ray observations that have higher signal--to--noise and
resolve some absorption feature or edge will allow more accurate models
for the absorption at $z_a$ = 0.0 and $z_a = 0.524$ toward AO 0235+164.

\citet{turnshek2003} also analyze two other DLA systems using Chandra X--ray Observatory data.
One of the systems, the DLA at $z_a = 0.313$ toward PKS 1127$-$145, was
analyzed by \citet{bechtold2001} as discussed above.
\citet{turnshek2003} find as did \citet{bechtold2001} that only an upper limit to the
metallicity can be established for this DLA system
although the details of the analysis are different.
\citet{turnshek2003} conclude that there is no evidence for a relatively high
metallicity in the PKS 1127$-$145 DLA system.
But they also find that even a zero-metallicity DLA
model does not fit the observed X--ray spectrum well so the metallicity is not well
determined.
\citet{turnshek2003} find that the $z_a = 0.394$ DLA system
toward S4 0248+430 could be metal enhanced with
an enhancement similar to the metallicity found in the $z_a = 0.524$ system toward
AO 0235+164.  But as noted by \citet{turnshek2003}, the S4 0248+430 line
of sight could have another high column density absorber at $z_a = 0.051$ which
could provide substantial X--ray absorption.
The likelyhood that the $z_a = 0.051$ is a DLA system is high because the spectrum
of S4 0248+430 shows extremely
strong Ca II and Na I absorption at $z_a = 0.051$ \citep{womble1990}.
The $N(H I)$ column at
$z_a = 0.394$ toward S4 0248+430, as discussed in \citet{turnshek2003},
is based not on a DLA measurement,
but on H I 21 cm observations and
an assumed spin temperature which increases the uncertainty in the metallicity.
The $z_a = 0.524$ DLA system toward AO 0235+164 provides the most
certain metallicity measurement of the three
DLA systems considered by \citet{turnshek2003}.
The other two DLA systems suggest that there is quite a range of metallicity
in the low redshift DLA systems.

\subsection{The 2175~\AA \ Dust Feature, Dust--to--Gas, and Dust--to--Metals}

The continuum spectrum of AO 0235+164 (Figure 1) shows a strong, broad
absorption feature centered at about 3300~\AA \ (observed), near the expected
wavelength of the 2175~\AA \ dust feature at $z_a$ = 0.524.
In order to analyze this feature, it is necessary to model the underlying
spectrum of AO 0235+164 and to have a numerical model for dust extinction.
The underlying spectrum of AO 0235+164 is assumed to be a power--law
continuum (or several power--laws connected at ``break--point'' wavelengths).
The numerical model includes narrow absorption features
in the UV (since these were necessary for the DLA analysis).
Narrow gaps in wavelength are excluded from the fit
at the locations of absorption features in the optical spectrum.
The dust extinction is modeled using the extinction formulae from
\citet[hereafter CCM]{cardelli1989} for Galactic dust
and the extinction formulae from \citet{pei1992} for LMC dust, SMC dust, and
a modified Galactic version (see below).
The advantage of the CCM formulae is that a larger range of Galactic
extinction curves can be fit using $R_V$ = $A_V$ / $E(B - V)$ as a variable
as well as $E(B - V)$.
The \citet{pei1992} extinction formulae fix $R_V$ at 3.1 for Galactic extinction
(the most common value found).
A summary of the fits to the spectrum of AO 0235+164 including dust extinction
is given in Table 1.
Figure 5 shows the continua from three $\chi^2$ fits to the
AO 0235+164 spectrum over the spectral region from 2500~\AA \ to 8700~\AA .
To emphasize the continuum features, only the 
continua are plotted above the spectrum (i.e. no absorption features are plotted
above the spectrum).
The fits shown are for models with Galactic, LMC, and SMC type dust
(fits 1, 2, and 3 respectively in Table 1).
Overlaid on the observed spectrum of AO 0235+164 in Figure 5 is the Galactic dust
model.
The Galactic dust model is the best fit of the three shown.

Multiple dust components are included in all of the fits, one at $z_a$ = 0
and one or more at $z_a$ = 0.524.
The $z_a$ = 0
Galactic absorption is fixed, for all the models listed in Table 1, with
$E(B - V)$ = 0.154 (based on the H I 21 cm as described above)
and $R_V$ = $A_V$ / $E(B - V)$ = 3.1.
The CCM formulae are used to model the dust at $z_a$ = 0.0.
As described in the previous section, the $E(B - V)$ at $z$ = 0.0
is not well determined by these spectra.
A range for the $z_a$ = 0.0 dust extinction can be estimated by varying $E(B - V)$ and
comparing, by eye, the depth of the resulting 2175~\AA \ feature to the observed data.
The range of values found is from $E(B - V)$ = 0.12 to $E(B - V)$ = 0.19 and this
range is used later to estimate part of the systematic errors in the
parameters to fit 1 (Table 1) which is our best estimate for the dust parameters
at $z_a$ = 0.524 along the AO 0235+164 line--of--sight. 

\placefigure{fig5}

Galactic type dust extinction at $z_a$ = 0.524 produces the best fit 
to the AO 0235+164 spectrum when comparing Galactic, SMC,
and LMC type dust.
The observed spectrum does not fit the CCM 
extinction function perfectly:  systematic deviations can be seen 
at $\lambda$ $>$ 8700~\AA , at $\lambda$ $<$ 2500~\AA , and 
on the red half of the $z_a$ = 0.524 2175~\AA \ feature (Figure 5).
The Galactic 2175~\AA \ feature varies in shape from line--of--sight
to line--of--sight \citetext{e.g. \citealp{fitzpatrick1986}},
so small deviations around 2175~\AA \ at $z_a$ = 0.524 are
not surprising.
The best fit parameters from fit 1 (Galactic dust)
are $E(B - V)$ = 0.227 $\pm$ 0.003 and $R_V$ = 2.51 
$\pm$ 0.03.  The above errors are only the formal errors in the $\chi^2$ 
fits.  We can estimate that part of
the systematic error due to uncertainties in the $z_a$ = 0.0 extinction
by using the above mentioned range of $E(B - V)$ = 0.12 to
$E(B - V)$ = 0.19 at $z_a$ = 0.0.
Re--fitting the observations with these values gives
$E(B - V)$ = 0.23 $\pm$ 0.01 and $R_V$ = 2.5 $\pm$ 0.2
at $z_a$ = 0.524.  Other systematic errors are likely because 
the extinction at $z_a$ = 0.524 is not a perfect match to ``typical'' 
Galactic extinction and does deviate significantly in the far--UV.
The best fit background source power law over 2500~\AA \ to 
8700~\AA \ has $\alpha$ = $-$1.75 $\pm$ 0.01 (formal error) 
for $f_{\nu} \propto \nu^{\alpha}$.  The systematic error due to 
the uncertain $z_a$ = 0.0 extinction increases the error and gives 
$\alpha$ = $-$1.75 $\pm$ 0.15.  The above formal errors from the 
$\chi^2$ fits and those listed
in Table 1 are probably severe underestimates of the true errors but 
they are an indication of what the statistical errors would be
if the shape of the extinction curve at $z_a$ = 0.524 and the
extinction parameters at $z_a$ = 0.0 were known accurately.

The extinction curves for the SMC and LMC differ greatly from the
extinction curves typically observed in the Galaxy.
The LMC extinction curve has a less pronounced
2175~\AA \ absorption feature compared to the Galaxy for
lines--of--sight with similar total $A_V$.
The SMC extinction curve has practically no 2175~\AA \ graphite
feature.
Figure 5 shows $\chi^{2}$ fits to
the observed AO 0235+164 spectrum using the SMC and LMC extinction 
curves from \citet{pei1992} at $z_a$ = 0.524.
As expected, the SMC type 
extinction does not fit the observations.
The LMC extinction 
model for the $z_a$ = 0.524 system cannot fit the observations
over the 2500~\AA \ to 
8700~\AA \ spectral region because the predicted UV extinction is
too strong when the model 2175~\AA \ feature has the same depth 
as the observed feature.
The best $\chi^{2}$ fit misses both in the continuum and in the depth
of the 2175~\AA \ feature.
In order to further explore the possibility of fitting the AO 0235+164
spectrum using an LMC type extinction curve, a model (fit 4 in Table 1)
was found that reproduces the 2175~\AA \ feature reasonably well.
But this model, that only fits the spectrum over the a restricted wavelength
range from 2800~\AA \ to 6000~\AA , has a number of unusual features.
The underlying spectrum is found to rise in units of $\nu f_{\nu}$ and
the total extinction at 2500~\AA \ (observed) is more than a
factor of 100 compared to a factor of 16 for the Galactic extinction model.
The unabsorbed underlying flux of 
$\sim$ 3 $\times$ 10$^{-14}$ erg cm$^{-2}$ s$^{-1}$ at $\sim$ 2500~\AA \ (at 
the time of these observations) - would make 
the de-reddened AO 0235+164 a very unusual BL Lac object relative 
to other strongly beamed and presumably less extincted BL Lac objects.
A simpler explanation is that the extinction in the $z_a$ = 0.524 DLA 
system is more similar to Galactic extinction than LMC extinction.

In order to examine the possibility of a less restrictive dust model,
a fit to the observations was made using a ``stacked'' absorber model at $z_a$ = 0.524.
One component consisted of a Galactic absorber and the other an LMC type absorber.
This fit is listed as number 6 in Table 1.
The result of the $\chi^{2}$ fit was to minimize the LMC
component ($E(B - V) \le 0.005$).
This shows that mixing LMC and Galactic type absorption does not improve
the fit to this absorption system.
A complete investigation of the kinds of extinction laws allowed by these
data is beyond the scope of this paper.
But amongst the Galactic, LMC, and SMC type extinction models, the Galactic type
absorption fits the data the best.
Galactic type absorption does fail to match the data below 2500 \AA \ (observed), but
this failure in the far--UV cannot be fixed by adding SMC or LMC type absorption to
create a multi-component absorber.

The most significant fractional deviations in Figure 5
between the observed spectrum and the
Galactic extinction model for the $z_a$ = 0.524 system
occur at $\lambda$ $\le$ 2500~\AA .  Attempts to fit a more
complete wavelength region, from 1690~\AA \ to 8700~\AA , using
a single power law and variable Galactic absorption
at $z_a$ = 0.524 result in a fit similar to that shown in Figure 5.
The furthest UV part clearly does not match the model spectrum
and the $\chi^2$ fitting process does not give the far--UV
points much weight because the S/N there is lower relative to 
the S/N further to the red.
This relatively strong observed far--UV spectrum (below 
about 2500~\AA \ observed) can result from either an intrinsic 
AO 0235+164 spectrum ($z_e$ = 0.94) that rises to the blue above 
the best fit power law or from an extinction curve at $z_a$ = 0.524 
that has less UV extinction than the CCM Galactic model.
Galactic extinction is known to vary from one line--of--sight to
another particularly in the far--UV, presumably because the number
of very small particles relative to other dust particles is highly
non--uniform \citetext{e.g. \citealp{fitzpatrick1999}}.
Using the \citet{pei1992} model for Galactic extinction and
reducing the coefficient for the far--UV extinction (to 0.73 of
the Pei value) produces a better fit to the overall spectrum (from 
1690~\AA \ to 8700~\AA \ observed).
This model is listed as fit 5 in Table 1.  However, this model does not match 
the data in detail (i.e. at $\le$ 5\% fractional variations everywhere).
A better match to the observed spectrum can be obtained by
using two power laws with $f_{\nu}$ $\propto$ $\nu^{-1.75}$ at
$\lambda$ $>$ 2549~\AA \ and $f_{\nu}$ $\propto$ $\nu^{+0.28}$ at 
$\lambda$ $<$ 2549~\AA \ (observed).
This model is listed as fit 7 in Table 1.
A steeply rising underlying spectrum 
to shorter wavelengths would miss the observed X--ray spectrum unless 
there is a further inflection at wavelengths between the observable 
optical--UV and X--ray spectral regions.  A complicated underlying 
spectrum with four power--law components or two power--law components 
plus a ``bump'' cannot be ruled out by the data.  A Galactic type 
extinction curve can describe the $z_a$ = 0.524 system, but it requires 
somewhat less than normal far--UV ($\lambda_{rest} < $ 1700~\AA ) 
extinction and minor deviations from the CCM extinction shape:  this 
can fit the observation, and be consistent with an underlying spectrum 
with one UV--optical power--law and one X--ray power--law component.
Without additional assumptions or constraints, it is not possible to 
determine conclusively both the underlying spectral shape and
the extinction curve.  For the remainder of the paper, we will 
primarily discuss the data in terms of the Galactic extinction model 
at $z_a$ = 0.524 with less far--UV extinction and a single power law 
intrinsic spectrum in the optical--UV spectral region.
But other combinations of extinction and underlying spectra can provide 
possible fits to these observations. 

The dust--to--gas ratio for DLA systems is often
given by a dimensionless parameter $k$ which is a ratio
between dust optical depth and $N(H I)$ defined by
$k \equiv 10^{21}(\tau_B/N(H I))$ cm$^{-2}$ \citep{pei1992}.
Using $E(B - V)$ = 0.23 $\pm 0.01$, $R_V$ = 2.5
$\pm 0.2$, and $N(H I)$ = 5 $\pm$ 1 $\times$ 10$^{21}$
gives $\tau_B$ = 0.74 and $k$ = 0.15 $\pm 0.03$.
The value of $k$ = 0.15 $\pm$ 0.03 is 0.19 $\pm$ 0.04 of the average Galactic
value of $k$ = 0.78 and similar to $k$ = 0.16 for the LMC \citep{pei1992}.
The observed $E(B - V)$ vs. $N(H I)$ correlations for the Galaxy, the 
LMC, and the SMC show considerable scatter from line--of--sight to 
line--of--sight, but the AO 0235+164 $z_a$ = 0.524 $N(H I)$ and 
$E(B - V)$ point is typical of the LMC values and well outside the scattered
Galactic values \citep{pei1992}.  Using the metallicity estimate 
from the previous section gives a dust--to--metals ratio, $k/Z$, of 0.27 
$\pm$ 0.12 of $(k/Z)_{Galactic}$.  For $d_m$, the dust--to--metals mass
ratio, assuming $d_m(AO)$ = 0.27 $\times$ $d_m(GAL)$, $d_m(GAL)$ = 0.51 
\citep{pei1999}, gives $d_m(AO)$ = 0.14 $\pm$ 0.06.  Assuming that $d_m(AO)$ 
scales from the Galactic as $ (k/Z)_{AO} / (k/Z)_{Galactic}$ implies 
that the density of dust grains is the same in the AO $z_a$ = 0.524 
system as it is in the Galaxy.  This assumption depends on the mix of 
grains because silicate and graphite grains have different densities.
With $d_m(AO)$ = 0.14, the $z_a$ = 0.524 DLA system has converted less
metals to grains compared to the Galaxy, LMC, or SMC which have
$d_m$ values of 0.51, 0.36, and 0.46 respectively.  Averaged over all 
objects as a function of redshift, $d_m$, the global dust--to--metals mass 
ratio, relates the comoving density of interstellar dust $\Omega_d$
to the comoving density of heavy elements $\Omega_m$ (all in units of the 
closure density) via $\Omega_d = d_m \Omega_m = d_m Z \Omega_g$
\citep{pei1999}.

\subsection{Diffuse Interstellar Bands}

The shortest wavelength and strongest of the known DIBs at 4428~\AA \ (rest)
appears in the Keck LRIS spectrum of AO 0235$+$164 at the expected 
wavelength, $\lambda_{obs} = 6747$~\AA , for $z_a$ = 0.524 (Figure 6).
The central depth, $A_c$, of the observed feature is 0.040 $\pm$ 0.005,
the equivalent width is 1.13~\AA \ $\pm$ 0.04~\AA \ (observed), and the 
width is about 32~\AA \ FWHM (observed).  The Keck LRIS spectrum is 
noisier further to the red due to incomplete correction for fringing, 
and thus the other DIBs are not cleanly detected but there is a depression 
at the predicted position of the 4882~\AA \ DIB.  The 
$z_a$ = 0.524 4428~\AA \ DIB also appears in the STIS CCD G750L spectrum
(Figure 1) at lower signal--to--noise.  The 4428~\AA \ DIB parameters for 
HD 183143, a line--of--sight commonly used for Galactic DIB comparisons,
are $W_{\lambda}$ = 3.4~\AA \ , $A_c$ = 0.15, and  FWHM = 18~\AA \
\citep{herbig1995}.  A comparison to the observed 4428~\AA \ DIB parameters 
at $z_a$ = 0.524 gives in rest units $W_{\lambda}(AO)$ / 
$W_{\lambda}(HD)$ = 0.22, $A_c(AO) / A_c(HD)$ = 0.26, and
FWHM(AO) / FWHM(HD) = 1.2.  The strength of the AO 0235+164
4428~\AA \ DIB is about 0.25 of the DIB observed toward
HD 183143 and the other, weaker DIBs are expected to roughly scale
by this factor of 0.25 in both central depth and $W_{\lambda}$
\citetext{e.g. \citealp{herbig1995}}.

\placefigure{fig6}

The presence of the DIBs in stellar spectra is correlated with
reddening by dust. 
The central depth, $A_c$, of the 4428~\AA \ DIB feature 
is roughly proportional to $E(B - V)$ in the Milky Way.
For a sample of Perseus OB1 stars, \citet{krelowski1987} 
find $A_c$ = 0.0117 + 0.0788 $E(B - V)$ (where $A_c$ is
expressed in fractional units).  For the observed $E(B - V)$ = 0.23 $\pm$ 0.01
in the $z_a$ = 0.524 AO 0235+164 system, the above equation gives
a predicted $A_{c}$ = 0.030 $\pm$ 0.002.  The observed $A_c$ is 0.040
with errors of about $\pm$ 0.005.  Given the errors in the observed
$A_c$ and $E(B - V)$ and the known scatter in 
the Galactic $E(B - V)$ -- $A_c$ correlation \citep{krelowski1987},
the DIB strength to reddening ratio in the $z_a$ = 0.524 DLA 
system is similar to the ratios found along Galactic lines--of--sight.

\section{Discussion}

The $z_a$ = 0.524 absorber toward AO 0235+164 shows dust absorption
with a 2175~\AA \ feature similar to that seen in the Galaxy.
The Galactic type extinction curves form roughly a one parameter family
parameterized by $R_V$ (CCM) with a range of observed
values 2.2 $\le$ $R_V$ $\le$ 5.8.  The best fit to the $z_a$ = 0.524 
extinction curve has $R_V$ = 2.5 $\pm$ 0.2, but the most common value 
of $R_V$ in the Galaxy, $R_V$ = 3.1 cannot be ruled out.  Fixing $R_V$ 
at 3.1 at $z_a$ = 0.524 and refitting the observations produces
a qualitatively similar fit to that shown in Figure 5.  Although formally 
a less acceptable fit, given the unknown nature of the deviations between 
the fit and the observations in the 1700~\AA \ to 2500~\AA \ (observed) 
wavelength region, it is reasonable to accept qualitatively similar 
extinction curves.  At values of $R_V$ $\ge$ 4.0, the CCM Galactic 
extinction curves redshifted to $z_a$ = 0.524 become a poor match 
qualitatively to the observed data.

If the extinction curve at $z_a$ = 0.524 were known to be exactly
the same as a typical Galactic curve, then it would be possible to
deredden the observed spectrum to obtain the underlying BL Lac
spectrum.  Figure 7 shows the result from dereddening the observed 
spectrum using the CCM model with $E(B - V)$ = 0.23 and $R_V$ = 2.5
at $z_a$ = 0.524 and $E(B - V)$ = 0.154 and $R_V$ = 3.1
at $z_a$ = 0.  These are the parameters from the $\chi^2$ fit described 
in \S3.3 (fit 1 in Table 1).  The UV continuum at $\lambda_{obs} \le$ 2500~\AA , rises to
shorter wavelengths.  The slope is steep with 
$\nu f_{\nu} \propto \nu^{+1.28}$ (as determined from fitting a two--piece 
power law with a break at 2549~\AA , fit 7 in Table 1).  Figure 8 shows the relationship 
of the dereddened AO 0235+164 spectrum in the optical--UV to the observed 
and modeled X-ray spectra.   The underlying best fit X-ray spectrum does 
have a positive slope in $\nu f_{\nu}$ units with 
$\nu f_{\nu} \propto \nu^{+0.22\pm0.09}$.  The spectra of BL Lac 
objects are thought to be produced by synchrotron emission at low 
frequencies (in the case of AO 0235+164 this produces the flux in the 
optical to UV frequency range falling in $\nu f_{\nu}$)
and a self Compton component at higher frequencies (rising in 
$\nu f_{\nu}$ to a cutoff frequency in the $\gamma$--ray range).
AO 0235+164 would have to have an additional component if the apparent
rising flux in the observed wavelength 
interval 1700~\AA \ $\le \lambda_{obs} \le$
2500~\AA \ were due to the underlying spectrum:  we consider this unlikely.  

\placefigure{fig7}

\placefigure{fig8}

An alternative to explaining the observed spectrum with Galactic
type absorption at $z_a$ = 0.524 -- which would require a relatively 
complex underlying spectrum -- is to allow for extinction laws
outside the ``typical'' Galactic.  As discussed in \S3.3, the 
\citet{pei1992} parameterization of the Galactic extinction law can be 
modified to produce less far--UV extinction by reducing the value of the
coefficient of that component.  It is not clear that this produces 
a physically reasonable extinction curve;  besides, the extinction curve 
that is produced does not fit the observed flux in detail under an assumption 
that the underlying flux is a single power--law component.
However, given the fact that the standard Galactic (CCM or \citealp{pei1992})
extinction curve does not fit most Galactic observations in
detail around the 2175~\AA \ feature and in the far--UV,
the observed spectra could result from Galactic type extinction
at $z_a$ = 0.524 with less absorption in the far--UV band, 
at $\lambda_{rest}$ $<$ 1700~\AA .  The observed variations in the 
Galactic extinction in the far--UV \citep{fitzpatrick1999} do suggest 
that the relative number distribution of small grains is an 
independent parameter in dust extinction.  An attempt to invoke an 
LMC or SMC type extinction curves to explain the extinction at 
$z_a$ = 0.524 does not work because the observed 2175~\AA \ feature is 
so strong.  A completely new kind of extinction curve is also possible, 
but in that case, there are few constraints on any measurable quantities 
such as $E(B - V)$ at $z_a$ = 0.524, because neither the underlying spectral
slope nor the unextincted flux are known.  

The absorber responsible for the $z_a$ = 0.524 absorption toward AO 0235+164
could be a galaxy about 2 arcsec south (object A) \citep{smith1977} or
another object (object A1) about 1 arcsec East \citep{yanny1989}.
Object A has a Seyfert 1 galaxy emission line spectrum with strong
associated absorption or weak BAL type absorption \citep{burbidge1996}.
Object A1 has [O II] $\lambda$ 3727 emission at $z_e$ = 0.524 based on
narrow band imaging with a filter tuned to redshifted $\lambda$3727
at $z$ = 0.524 \citep{yanny1989}.  Spectroscopy of object A1 is needed 
to verify that it is indeed a galaxy at $z$ = 0.524, but the [O II] results 
make that likely.  Although the AO 0235+164 -- subtracted HST WFPC2 image 
shown in \citet{burbidge1996} is noisy, the image suggests that both 
A and A1 are disk galaxies \citep{burbidge1996}.  With this, the existing 
imaging, spectroscopy of the nearby objects, and absorption spectra are 
all consistent with the interpretation that the $z_a$ = 0.524 absorption 
is produced by a line--of--sight through a galaxy disk or pair of galaxy 
disks.  The impact parameter is about 14 ${h_{65}}^{-1}$ kpc for object 
A and 7 ${h_{65}}^{-1}$ kpc for A1 assuming $\Omega_M = 0.3$,
$\Omega_{\Lambda} = 0.7$ and where $h_{65}$ = $H_0 / 65 $ km s$^{-1}$ 
Mpc$^{-1}$.  Therefore, depending on the inclination, the $z_a$ = 0.524 
DLA system may sample a galaxy disk at a distance from the nucleus which is 
not too dissimilar to the position of the Sun in the Milky Way.
It is also possible that a galaxy--galaxy interaction plays a part in
determining the properties of this absorber if A1 is at $z$ = 0.524 and 
interacting with A \citep{yanny1989}.  The 21 cm H I absorption has multiple 
deep components \citep{wolfe1982} and the observed Mg II absorption 
has a large total velocity width, 250 km s$^{-1}$, and shows more 
components than the 21 cm \citep{lanzetta1992}.  The large range of 
absorption velocities and the large number of individual absorption 
components might suggest absorption by an interacting
pair rather than absorption by a single isolated disk galaxy.
The low value of the dust--to--metals mass ratio $d_m$ for this system,
0.14 vs. 0.51 for the Galaxy, could be produced by the release of metals from
dust due to cloud--cloud collisions.
If the absorption is primarily produced by gas involved in an interaction
between objects A and A1, then the $z_a$ = 0.524 system may not be typical
of high metallicity DLA disk absorbers.

Object A may produce a DLA absorber that is not typical
of DLA absorbers because it has intrinsic
absorption and an AGN type mass outflow.
Because the geometry and extent of AGN mass outflows in general are not well known,
it is possible that a mass outflow could change the characteristics of
disk--like gas far from the central source.
As mentioned above, the impact parameter between object A and the BL Lac
object corresponds to 14 ${h_{65}}^{-1}$ kpc.
The observations of object A \citep{burbidge1996} are very low resolution and constitute
a single epoch so it is not possible to determine if the intrinsic
absorber is close to the AGN using partial covering and/or time variability.
Other examples of DLA systems associated with AGN are needed to determine
if there is any correlation between the AGN properties and the DLA absorption properties in
the host galaxies.

The dusty $z_a$ = 0.524 DLA system toward AO 0235+164 is an example of the
kind of DLA system that will suffer selection effects \citep{pei1995}
when the background sources are selected from a magnitude limited
sample.  The large inferred extinction is shown graphically in Figures 7 and
8 where the dereddened and observed spectra are both plotted.
At the observed frame B band, 4400~\AA , the extinction due to dust at 
$z_a$ = 0.524 is 1.3 magnitudes ($R_V$ = 2.5 and $E(B - V)$ = 0.23).
The observed apparent number counts of quasars as a function of apparent
magnitude, $m$, can be described as $N(m) \propto 10^{0.4\beta m}$
where $N(m)$ is the number of quasars in units of deg$^{-2}$ mag$^{-1}$.
For B band counts \citet{pei1999} find $\beta$ = 2.0.
This fairly steep rise in the number of quasars as a function of $m$ holds
for $B$ $\le$ 20. Because the number counts are an exponential, the 
fraction of quasars missed due to dust obscuration is independent of the 
limiting magnitude of the survey for $B_{lim}$ $\le$ 20.  Integrating $N(m)$ 
with $\beta$ = 2.0, the fraction of DLA systems missed 
due to selection effects is  $1 - e^{-1.84 A_{B(obs)}}$
where $A_{B(obs)}$ is the extinction of the DLA system at
the observed frame B band.
Systems similar to the $z_a$ = 0.524 absorber, with $A_{B(obs)}$ = 1.3,
would be underrepresented by a large factor with
91\% missing from a B magnitude limited quasar survey.
If there were systems like this at a redshift of $z_a$ = 2.0,
the extinction at 4400~\AA \ (obs) would be about 2.0 magnitudes and
about 97\% would be missing due to selection effects.

Absorption systems, with $N(H I)$ $\sim$ 5 $\times$ 10$^{21}$
cm$^{-2}$, are near the high $N(H I)$ turnover
(high $N(H I)$ knee) in the H I column density distribution
\citep{storrie-lombardi1996a,rao2000}.  It is important to count 
such systems accurately in order to measure the evolution with $z$ of 
the mean contribution of neutral gas to the cosmological mass density, 
$\langle\Omega_g\rangle$ \citep{storrie-lombardi1996b}.
The observed H I column density distribution for $N(H I)$ below the 
high $N(H I)$ knee is typically described as a power law with 
$f(N_H) = f_{*}{N_H}^{-\beta}$.  Because the observed value of 
$\beta$ is around 1.5 -- 1.7 \citep{storrie-lombardi1996a} and one is 
integrating $N f(N) dN$, the highest column density systems, up to the 
high $N(H I)$ knee, contribute the most to the integrated $N(H I)$ at 
each redshift.  Absorption systems that are missed due to dust can 
contribute significantly to $N(H I)$ without contributing much added 
search path thus leading to an underestimate of $\langle\Omega_g\rangle$
which is often estimated
from the observed function, $\Omega_{DLA}(z)$,
which just includes the DLA contribution.

At low redshift, the observed DLA H I column density distribution shows 
a relative excess of high column density systems compared to the high redshift
H I column density distribution \citep{rao2000}.
Rao and Turnshek find no observed evolution of $\Omega_{DLA}(z)$ with $z$.
\citet{rao2000} consider the AO 0235+164 $z_a$ = 0.524 DLA system to 
be a biased line--of--sight because it has a 21 cm H I detection, 
but the 21 cm H I observations of \citet{roberts1976} were tuned 
to $z_a$ = 0.524 based on the detection of Mg II absorption at that 
redshift.  Thus the $z_a$ = 0.524 system could be considered a Mg II selected
absorption system and is not a system discovered by 21 cm H I absorption.
The OI 363 $z_a$ = 0.0912 and $z_a$ = 0.2212 absorption systems,
both found to have DLA absorption by \citet{rao1998},
also have 21 cm H I absorption \citep{lane1998}.  The OI 363 absorption 
systems are included in the Rao and Turnshek survey statistics and the 
AO 0235+164 $z_a$ = 0.524 system is not included.  The point of the 
above discussion is that the sample selection criterion used by 
\citet{rao2000}, namely the exclusion of all previously known 21 cm 
absorbers, is not a completely robust criterion for rejecting the Mg II
selected absorber toward AO 0235+164.  The ``no known 21 cm systems'' rule 
was used to avoid a bias in favor of high $N(H I)$ systems, but, in a 
few cases, that rule may produce a bias against high $N(H I)$ systems.
Maybe the $z_a$ = 0.524 absorber would have made the \citet{rao2000}
sample if nothing were known about the 21 cm absorption
or maybe the $z_a$ = 0.524 absorber without dust would have made that sample.
If the AO 0235+164 $z_a$ = 0.524 absorber $N(H I)$ were included with the
16 other DLA systems in Rao and Turnshek, it would increase the total
$N(H I)$ in that low $z$ sample by 22\% .

\citet{rao2000} use their low redshift sample of DLA systems to determine the
low redshift end of $\Omega_{DLA}(z)$.
Their sample is binned into two bins with the lowest redshift bin including objects with
redshifts from $z$ = 0.0 to 0.8.
If AO 0235+164 with $z_a$ = 0.524 were to be included in this bin,
the total column density in this bin increases by
about 35\% while the search interval increases only slightly.
The $log_{10}\Omega_{DLA}(z)$ value in this bin would increase by about 0.13 from
-2.65 to -2.52 for the $q_0 = 0.5$ model plotted in \citet{rao2000}.
If the inclusion of AO 0235+164 were the only change to the $\Omega_{DLA}(z)$
function illustrated by \citet{rao2000}, the shape of the function would suggest
an increasing $\Omega_{DLA}(z)$ to lower redshifts - but the error bars in the
values of the binned $\Omega_{DLA}(z)$ are too large make any definite statements.
This does illustrate the importance of high $N(H I)$ systems to the determination
of $\Omega_{DLA}(z)$, and missing a few such systems due to dust
could change the observationally determined evolution of 
$\Omega_{DLA}(z)$ if dust evolves with redshift.

A number of studies use DLA and dust characteristics to estimate the 
corrections for selection effects.  These studies include \citet{fall1993},
\citet{pei1995}, and \citet{pei1999}.  The $z_a$ = 0.524 absorption 
system toward AO 0235+164 has some dust characteristics that differ 
from the assumed dust properties in \citet{pei1995} and \citet{pei1999}.
Pei and Fall (1995) \nocite{pei1995} assume that the extinction 
curve for DLA dust is intermediate between the LMC and SMC extinctions 
at $\lambda$ $<$ $2000$~\AA \ while the $z_a$ = 0.524 absorber resembles 
more closely the Galactic extinction curve, but with possibly less 
extinction than Galactic at $\lambda$ $<$ $1700$~\AA .  The LMC and SMC 
extinction curves have more extinction at $\lambda$ $<$ 1700~\AA \ than
the Galactic extinction curve.  The $z_a$ = 0.524 system has a low dust 
to metals ratio $k/Z$ of 0.27 $\pm$ 0.12 ($k/Z$)$_{Galactic}$ while
\citet{pei1995} and \citet{pei1999} assume that dust--to--metals is 
roughly constant at the Galactic value.  A lower dust--to--metals and 
less UV extinction would produce less correction for dusty DLA systems
than the corrections found by \citet{pei1995}.  The observation of one 
dusty low redshift DLA system with Galactic type dust toward a high 
apparent brightness, radio selected BL Lac object, a rare type of AGN, 
suggests that dusty systems with these dust characteristics may be common.
The dust extinction in DLAs may be more complex than the model
of \citet{pei1995} with SMC or LMC type dust typical in low metallicity
systems and Galactic type dust common in higher metallicity systems.
More dusty DLA systems must be observed before the corrections
for DLA systems missed due to dust extinction can be considered accurate.

Corrections for dust are important in the determination
of the $N(H I)$ weighted mean metallicity [$\langle Z(z) \rangle$]
in DLA systems.  \citet{prantzos2000} calculate a model for absorption 
by disk systems which shows that the observed lack of metallicity 
evolution in DLAs \citetext{e.g. \citealp{pettini1999}}
is a result of observational bias effects, in particular the effect of
dust extinction in high column density, high metallicity systems.
At the current time, because the number of well studied DLA systems is 
small, the inclusion or exclusion of individual systems with a high 
metallicity and a high $N(H I)$ can greatly influence the empirically 
determined [$\langle Z(z) \rangle$] \citep{prochaska2000}.
\citet{pettini1999} find for the lowest redshift bin, $z$ $<$ 1.5,
that [$\langle$Zn/H$\rangle$] = $-$1.03 $\pm$ 0.23 (on a logarithmic scale 
relative to solar).  The \citet{pettini1999} measurement is based
on 10 DLAs with $\sum N(H I)$ = 7.7 $\times$ 10$^{21}$ cm$^{-2}$.
If the $z_a$ = 0.524 system, with $N(H I)$ = 5$\pm$1 $\times$ 10$^{21}$ 
cm$^{-2}$ and $Z$ = ($0.72 \pm 0.28$) $Z_{\sun}$, were included
in the bin with [Zn/H] = $\log_{10} (0.72) = -0.14$,
then the $N(H I)$ weighted average would increase
to [$\langle$Zn/H$\rangle$] = $-$0.47, i.e. the metallicity in this 
bin would jump from 0.093 $Z_{\sun}$ to 0.34 $Z_{\sun}$.

The AO 0235+164 $z_a$ = 0.524 DLA system cannot be included in the 
sample of \nocite{pettini1997b} Pettini et al.~(1997b, 1999):  this 
is because it does not have a metallicity determined from Zn II measurements.
The X--ray determined metallicity in the $z_a$ = 0.524 system is
weighted most heavily to the abundance of oxygen with some contribution
from carbon and iron (\S3.2).
Oxygen and zinc may not reflect the same metallicity so it would be
important to measure the Zn II lines in the AO 0235+164 spectrum.
The Zn II lines at 2025~\AA \ and 2062~\AA \ will appear at 3086~\AA \ and
and 3142~\AA \ at $z_a$ = 0.524 and, being near the atmospheric cutoff in
an AGN with a red observed spectrum, these lines will be difficult to observe.
\citet{pettini1997b} chose Zn as a metallicity indicator because
Zn is relatively resistant to depletion onto dust grains.
Measurements of metallicity using X--ray absorption are also
insensitive to metal depletion onto dust grains.
The Zn II based metallicities and the X--ray determined metallicities will
differ if the abundances reflect $\alpha$ process enhancements and the
X--ray analysis is done assuming no $\alpha$ process enhancement.

\citet{turnshek2003} consider the evolution of metallicity with redshift
using three Chandra X--Ray Observatory determined metallicities (AO 0235+164 $z_a = 0.524$,
PKS 1127$-$145 $z_a = 0.313$, and S4 0248+430 $z_a = 0.394$) combined
with the \citet{pettini1999} Zn II based metallicities.
\citet{turnshek2003} consider the case for solar metallicities and alpha process
enhanced metallicities (by a factor of 2.5) in the absorber and find different
equivalent Zn II metallicity values in the two cases.
As described in \S3.2, \citet{turnshek2003}
find a metallicity about one third the value
in this paper for the $z_a$ = 0.524 DLA toward AO 0235+164
when models with similar assumptions are compared.
\citet{turnshek2003} find that the metallicities for the three DLA systems
might provide support for an increasing
[$\langle Z(z) \rangle$] to lower redshifts using the \citet{pettini1999} Zn based
values to determine the high redshift [$\langle Z(z) \rangle$] points.
The higher metallicity value found in this paper would tend to make the evolution
of [$\langle Z(z) \rangle$] toward lower $z$ more pronounced.
However, \citet{turnshek2003} conclude that the trend in the lowest redshift
points in their [$\langle Z(z) \rangle$] analysis depends on the treatment (exclusion
or inclusion) of the PKS 1127$-$145 DLA which has as large an H I column
density as the AO 0235+164 DLA and is treated as having zero metallicity.
Also \citet{turnshek2003} find that the lowest
redshift value of the [$\langle Z(z) \rangle$] function
depends strongly on the assumption of solar relative abundances versus $\alpha$ process
enhanced abundances.
The current samples of DLAs are just too small to determine
the low redshift end of [$\langle Z(z) \rangle$] function because most of the contribution
in $N(H I)$ comes from a few extremely high column density systems like the AO 0235+164 DLA.
The inclusion or exclusion of a single system can greatly influence the $N(H I)$
weighted quantities like [$\langle Z(z) \rangle$]: in particular the exclusion of
the AO 0235+164 DLA.
A large change in metallicity due to the
inclusion or exclusion of one system points out how fragile the current
[$\langle Z(z) \rangle$] measurement is due to small number statistics.
The difference in metallicity found in this paper vs. the values found by
\citet{turnshek2003} are also problematical in determining [$\langle Z(z) \rangle$].
An improved determination can be made using a simultaneous soft X--ray observation
and an HST observation to determine $N(H I)$ based on the DLA line.
This would reduce the uncertainty due to the possible time variability of the
absorption.
Unfortunately, AO 0235+164 is often too faint to be economically observed
with HST in the UV.
A high resolution observation of the Zn II lines at $z_a = 0.524$ would also be very
valuable, but difficult due to the atmospheric cutoff and the faintness of the
BL Lac object at these wavelengths.
 
At the present time, the value of [$\langle Z(z) \rangle$] at low redshift
is strongly influenced by the metallicity and by the inclusion or exclusion
in the low redshift sample
of one system, the $z_a = 0.524$ DLA toward AO 0235+164.
AO 0235+164 is now the prototypical line--of--sight
substantially reddened by dust extinction
due to an intervening DLA system.
This illustrates the potential importance of high
$N(H I)$ high $Z$ systems that are missed in magnitude limited
surveys due to dust extinction.  Observations of samples of quasars 
free from selection effects due to extinction will eventually resolve 
the issue of missed DLA systems with dust.  The CORALS survey 
\citep{ellison2001}, which selects background quasars for DLA systems
on the basis of radio emission only, has found a relatively
small correction (less than a factor of two) to the neutral gas fraction.
The question of the metallicity correction due to dust selection
effects remains unresolved because the high column density systems 
such as AO 0235+164 $z_a = 0.524$ are so rare that the number of systems
is poorly constrained by the current sample \citep{ellison2001}.

\section{Summary}

The properties of the $z_a$ = 0.524 absorption system
toward AO 0235+164 are intermediate between the high
redshift DLAs and the Milky Way Galactic disk.  
The absorber at $z_a$ = 0.524 has many qualitative similarities
in absorption to the Galactic disk around the 
position of our Sun.  The spin temperature at $T_s$ = 220$\pm$60 K 
is relatively low compared to other
DLAs with 21cm absorption and roughly near the typical value for the Galaxy.
There is dust extinction at $z_a$ = 0.524 with a 2175~\AA \ feature while
other DLAs do not show this feature.  The best fit to Galactic 
extinction over 2500~\AA \ $\le$ $\lambda_{obs}$ $\le$ 8700~\AA \ at
$z_a$ = 0.524 gives $E(B - V)$ = 0.23 $\pm$ 0.01 and $R_V$ = 2.5 $\pm$ 0.2.
The diffuse interstellar bands that are associated with dust are present,
and the DIB strength matches that predicted from the Galactic 
DIB vs. $E(B - V)$ correlation to within about 25\%.  
The metallicity of the system is relatively high 
compared to most DLA systems, $Z$ = $0.72 \pm 0.28$ $Z_{\sun}$.
Although it should be noted that an independent X--ray measurement
using a Chandra X--ray Observatory ASIS--S sepctrum gave (with
similar assumptions) $Z$ = $0.24 \pm 0.06$ $Z_{\sun}$ for the
$z_a = 0.524$ system toward AO 0235+164 \citep{turnshek2003}.
The difference amounts to about
2$\sigma$ and could reflect substantial time variability between our
Asca observations from 1998 February 11 and the Chandra X--Ray Observatory
observations on 2000 September 20.  
There are also some quantitative differences between the $z_a = 0.524$
DLA system toward AO 0235+164 and Galactic absorption.  The dust--to--gas ratio 
of $k = 0.15 \pm 0.03$ is more similar to that found in the LMC.
The dust--to--metals of $k/Z$ = 0.27 $\pm$ 0.12 $(k/Z)_{Galactic}$ is probably
not compatible with Galactic in spite of the large possible errors.
It is more similar the dust--to--metals of $k/Z$ = 0.5 $(k/Z)_{Galactic}$ 
estimated from element depletions in much lower metallicity high $z$
DLA systems \citep{pettini1997a}.  The dust at $z_a$ = 0.524 may also 
differ from Galactic dust in the shape of its extinction curve.
The dust either has less extinction than Galactic dust at UV wavelengths
shortward of $\sim$1700~\AA , or AO 0235+164 has a complex intrinsic 
spectrum with a steeply rising far--UV part (in $\nu f_{\nu}$ units)
at $\lambda$ $<$ 1300~\AA \ (rest $z_e$ = 0.94).

The $z_a$ = 0.524 absorption system is not dominated by SMC
or LMC like dust because the SMC extinction curve
would produce an un--detectable 2175~\AA \ feature
and the LMC extinction curve produces far too much UV extinction relative
to the 2175~\AA \ depth to be a good match to the observed spectrum.
Although the extinction due to the dust at $z_a$ = 0.524 is closer to 
Galactic than to that for SMC or LMC, substantial deviations from the
family of Galactic extinction curves cannot be ruled out.

The AO 0235+164 $z_a$ = 0.524 absorption system provides an excellent
illustration of the selection effects due to dust predicted by 
\citet{pei1995}.  The dust extinction at the observed B band is 
$\sim$1.3 mag, and would result in 91\% of such systems
being missed in a B magnitude limited quasar survey.
Dusty systems such as the AO 0235+164 $z_a$ = 0.524 absorber, with $N(H I)$ =
5 $\times$ 10$^{21}$ cm$^{-2}$, and $Z$ = 0.72 $Z_{\sun}$, might contribute a
significant fraction to the H I, dust, and metallicity of the Universe
as a function of redshift.  The dust--to--metals mass ratio of this 
one system is much less than the Galactic dust--to--metals ratio
assumed by \citet{pei1995} in calculating the correction for
DLA systems obscured by dust.  If the dust--to--metals ratio is 
consistently low in DLA systems, then, perhaps, dusty systems are less 
important than previously assumed.  On the other hand, if one in ten 
systems has both high metallicity and dust, these
systems can dominate the gas and dust phase metal inventory of the Universe.
The contribution of dusty systems is still an open question.  
These observations show that there are dusty systems with properties
intermediate between the low metallicity DLA systems that have been
studied previously and our Galaxy.

\acknowledgments

The authors wish to recognize and acknowledge the very significant
cultural role and reverence that the summit of Mauna Kea has
always had within the indigenous Hawaiin community.
We are most fortunate to have the opportunity to conduct observations
from this mountain.
Support for this work was provided by:  NASA through grant
number GO--7294 from the Space Telescope Science Institute, which
is operated by AURA, Inc., under NASA contract NAS5--26555;  through NASA 
Chandra program, via Smithsonian Institution grant GO1-2113X, and through 
Department of Energy contract no. DE-AC03-76SF00515 to the Stanford Linear 
Accelerator Center.

\clearpage
\figcaption[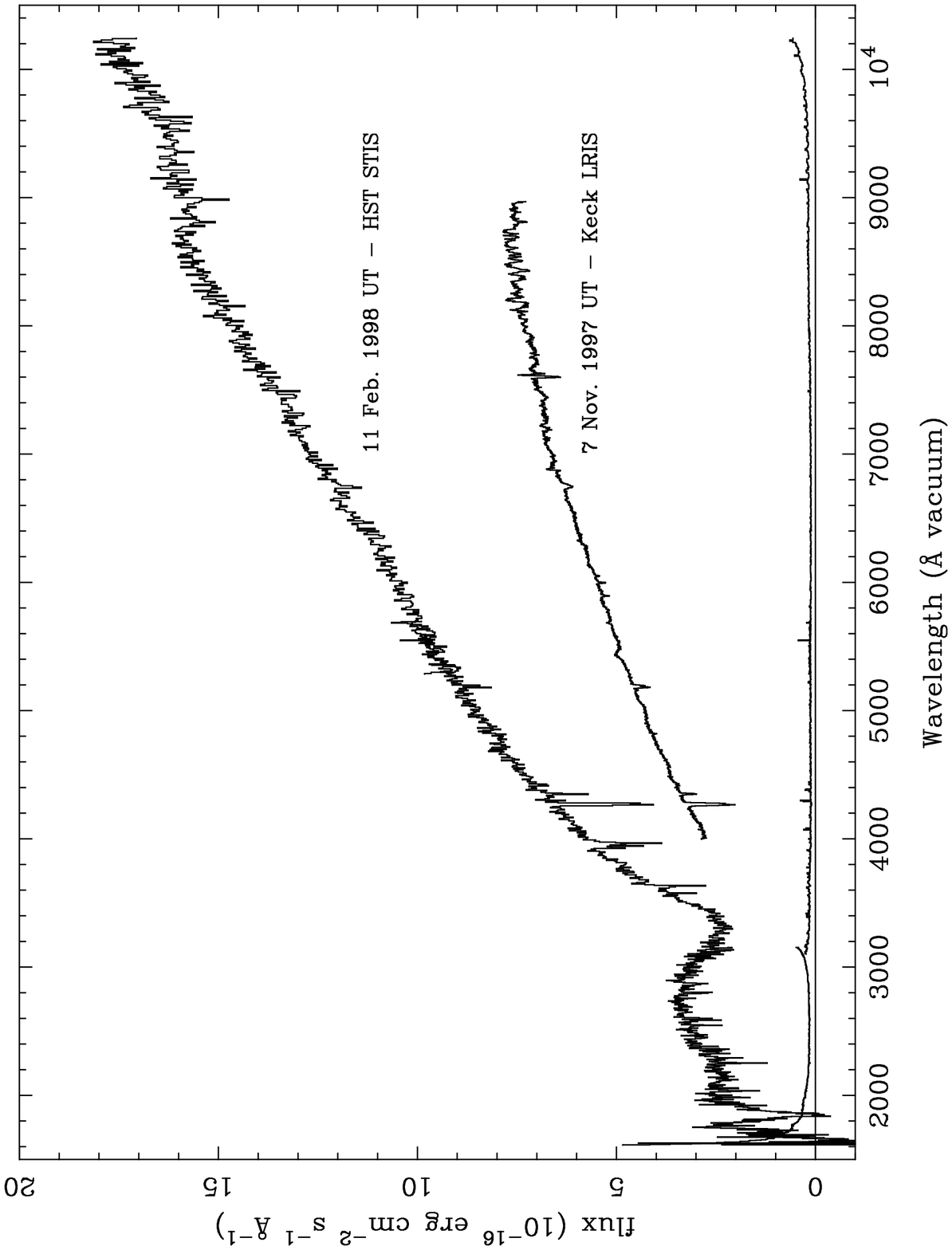]{Spectra of AO 0235+164.  The upper spectrum is the
STIS spectrum obtained 11 February 1998 (UT).  All three individual spectra,
NUV--MAMA G230L, CCD G430L, and CCD G750L, have been binned 2:1.
The lower spectrum shows the Keck LRIS spectrum from 7 November 1997 (UT).
The bottom trace shows the 1 $\sigma$ error in the unbinned STIS
spectrum \label{fig1}}

\figcaption[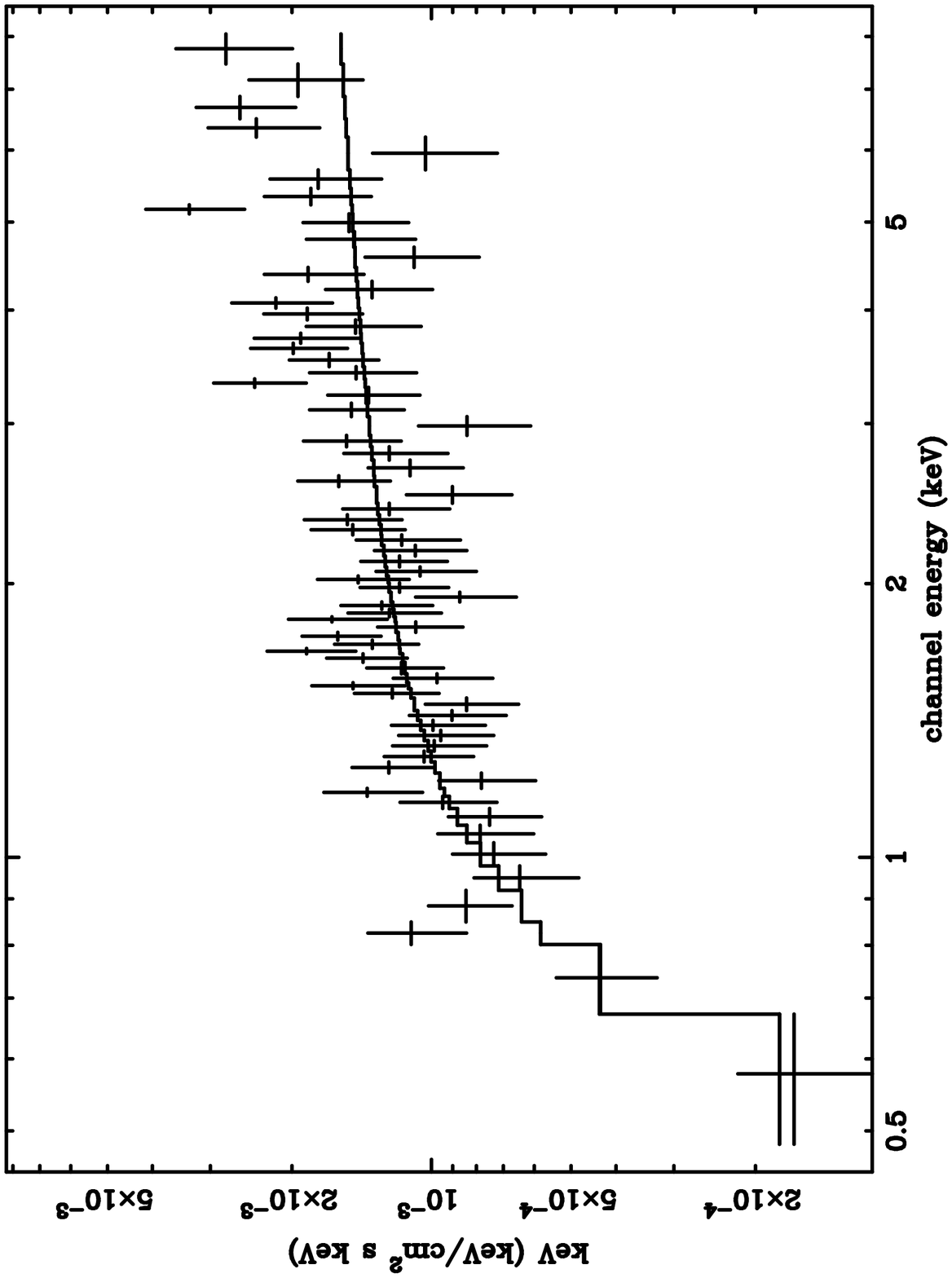]{The X--ray spectrum of AO 0235+164
from a representative Asca detector as observed on
February 11--12, 1998.
The spectrum as shown is corrected for the instrumental response.
Also shown is a fit using a power law spectrum with absorption
due to the Galaxy and the $z_a$ = 0.524 absorption system toward AO 0235+164.
\label{fig2}}

\figcaption[f3.eps]{Confidence regions ($\chi^{2}_{min}$ + 2.3, 4.6, and
9.2) for the absorption models fitted simultaneously to the ROSAT and 
1994 and 1998 Asca data.  The ordinates are the columns of 
the primordial absorber (consisting of H and primordial He) 
and the chemically processed absorber (no H, and only the 
processed portion of He, plus all other elements), both parametrized 
via hydrogen column density of material at solar abundances.  The 
STIS measurement yielding the column of H I of $5 \times 10^{21}$ cm$^{-2}$ 
implies the metallicity of $\sim 0.36/0.5 = 0.72$ solar \label{fig3}}

\figcaption[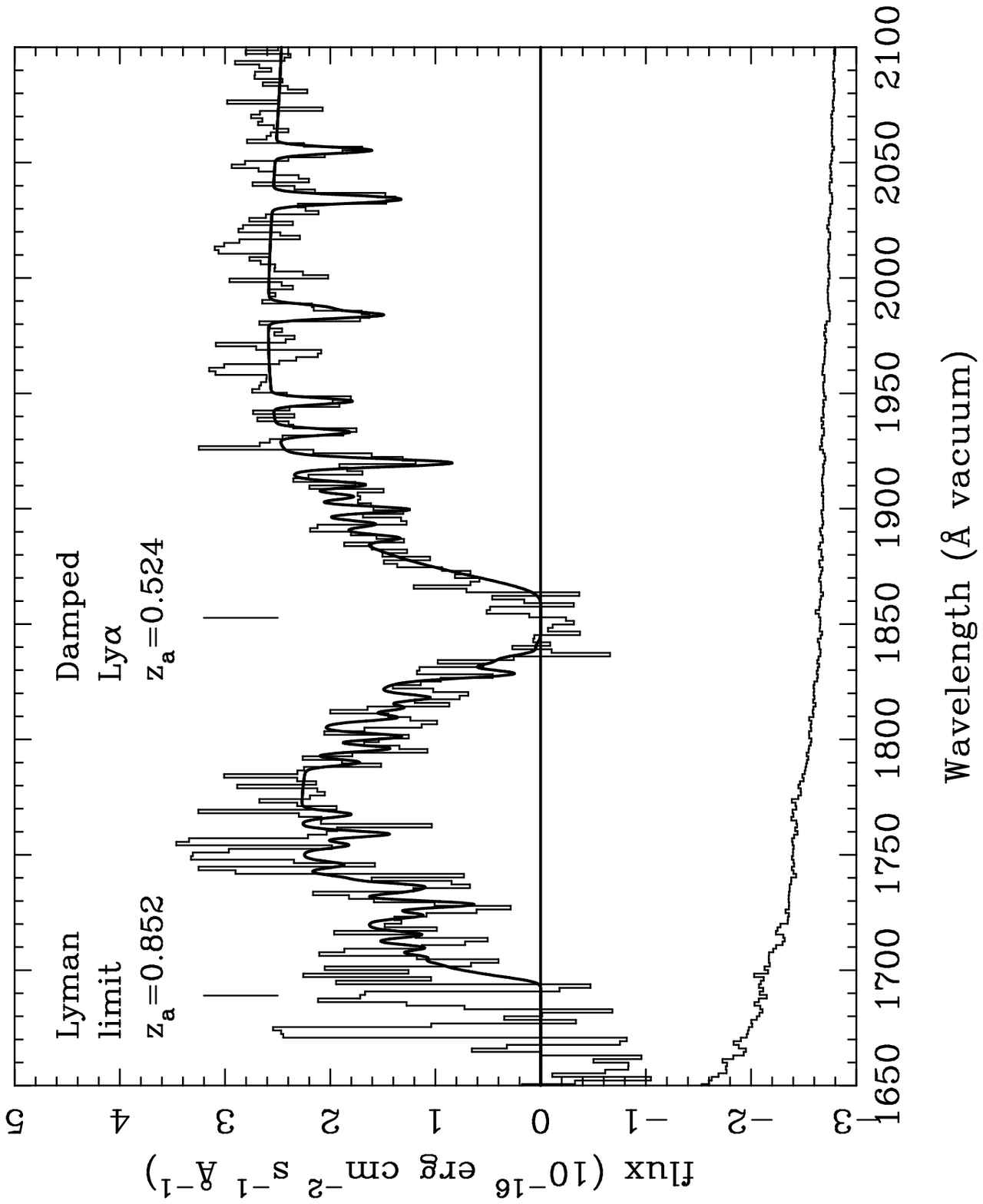]{Spectrum of AO 0235+164 around the $z_a$ = 0.524
DLA line.  The STIS NUV--MAMA G230L spectrum is shown as a histogram,
the one $\sigma$ errors are shown shifted below the spectrum, and
a model obtained from a $\chi^2$ fit to this spectral region with $N(H I)$ =
5 $\times$ 10$^{21}$ cm$^{-2}$ is shown as a heavy solid line superposed
on the data \label{fig4}}

\figcaption[f5.eps]{HST STIS spectra of AO 0235+164 from Figure 1
and Galactic, LMC, and SMC dust model $\chi^{2}$ fits.
The data are binned 2:1. 
For clarity, the continua of the fitting functions are shown
displaced above the data.
Superposed on the data is the Galactic model.
\label{fig5}}

\figcaption[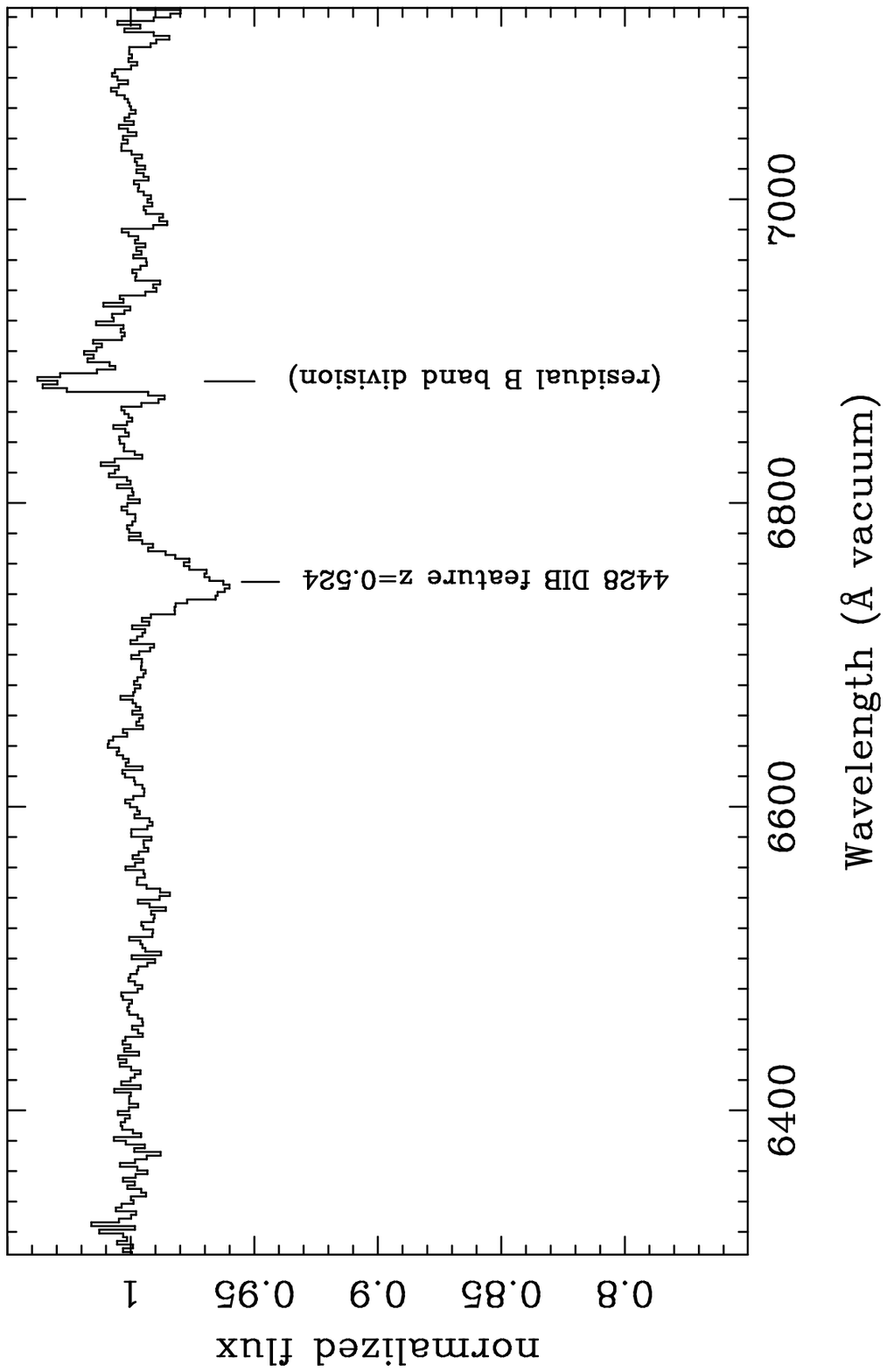]{The Keck LRIS spectrum of AO 0235+164 showing
the spectral region around the $z_a$=0.524 4428~\AA \ DIB feature.
The spectrum is plotted as normalized flux versus wavelength.
The zero point for normalized flux is below the lower left corner
of the plot in order to display the low contrast DIB feature.
\label{fig6}}

\figcaption[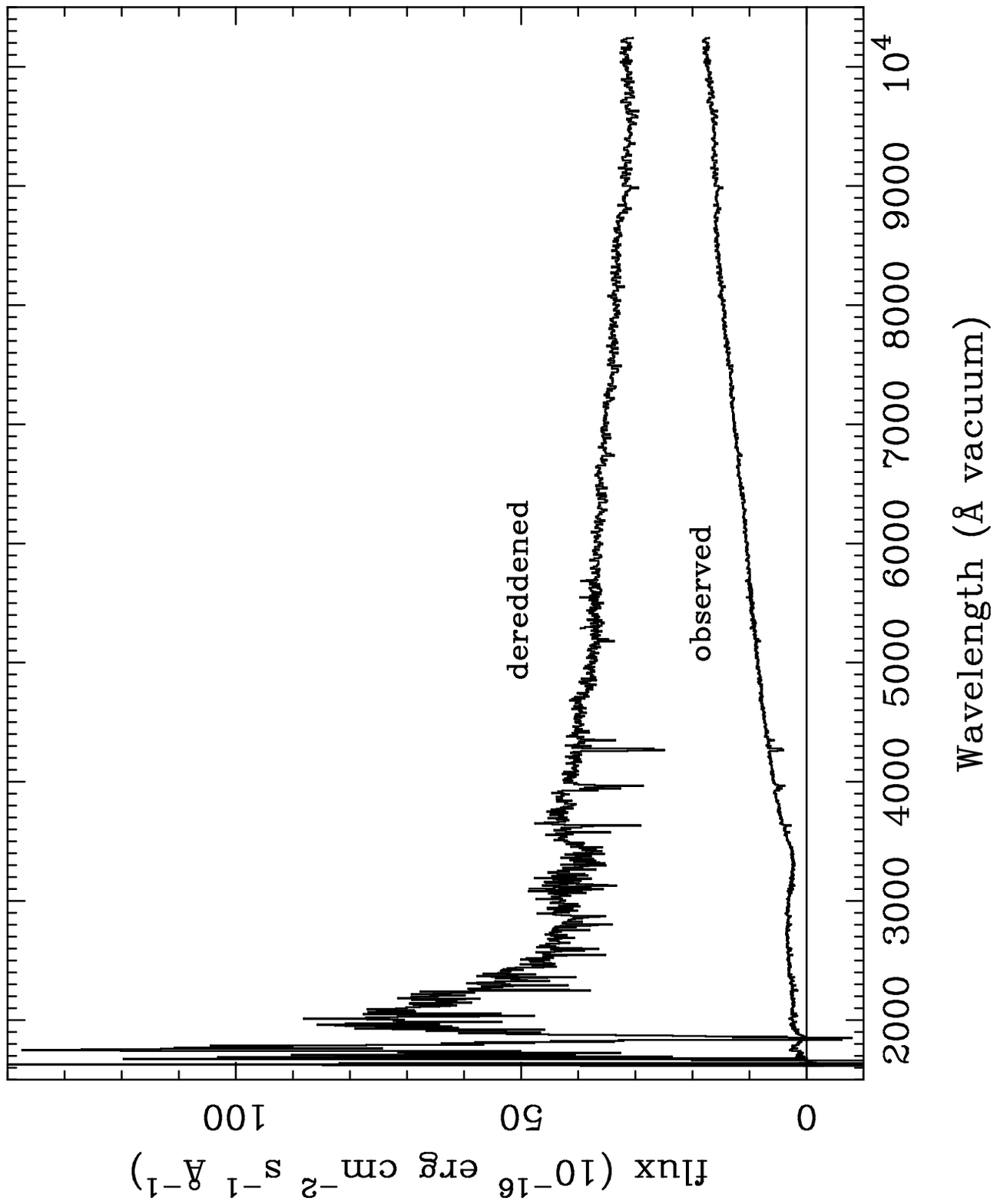]{The dereddened (upper) and observed (lower)
HST STIS spectra of AO 0235+164 shown as 
flux vs. wavelength.
The STIS NUV--MAMA data are binned 3:1 and the
STIS CCD data are binned 2:1.
\label{fig7}}

\figcaption[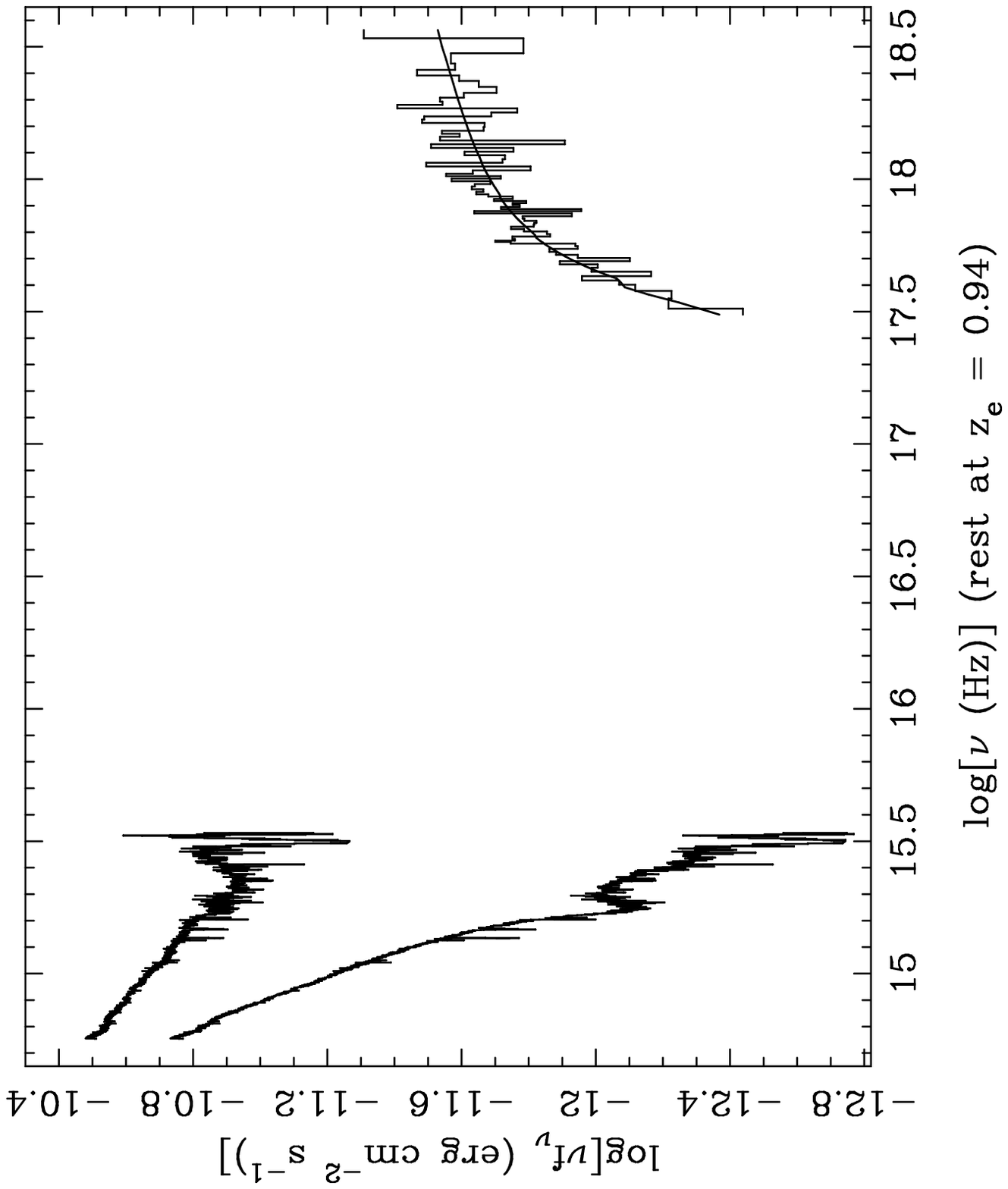]{The observed (lower) 
and dereddened (upper -- see text for
parameters) HST STIS spectra of
AO 0235+164 obtained February 11, 1998, shown as
$log_{10}(\nu f_{\nu})$ vs $log_{10}(\nu)$.
Also shown is the observed Asca X--ray spectrum from February
11--12, 1998.
The data have been shifted to the
rest frame of the $z_e$ = 0.94 BL Lac object.
The STIS NUV--MAMA data are binned 3:1, the
STIS CCD data are binned 2:1, and deep absorption
features have been arbitrarily clipped.
\label{fig8}}
\clearpage

\plotone{f1.eps}
\vskip 0.2cm
(Figure 1)
\clearpage
\epsscale{0.50} 
\plotone{f2.eps}
\vskip 0.2cm
(Figure 2)
\epsscale{0.95} 
\plotone{f3.eps}
\vskip 0.2cm
(Figure 3)
\plotone{f4.eps}
\vskip 0.2cm
(Figure 4)
\plotone{f5.eps}
\vskip 0.2cm
(Figure 5)
\plotone{f6.eps}
\vskip 0.2cm
(Figure 6)
\plotone{f7.eps}
\vskip 0.2cm
(Figure 7)
\plotone{f8.eps}
\vskip 0.2cm
(Figure 8)

\clearpage
\begin{table}
\centerline{\bf {TABLE 1}}
%\vspace{10pt}
\centerline{\bf{AO 0235+164 Spectral Fits with Dust Extinction}}
\vspace{1ex}
\begin{center}
\begin {tabular}{cccccccccccc}
%\small
\tableline
\tableline
% \vspace{1ex}
Fit & Dust Type & Ref.$^{\dagger}$ & E(B - V) & $\sigma^{\dagger \dagger}$ &  $R_V$   &
 $\sigma^{\dagger \dagger}$ & $\alpha^{\ddagger}$ & $\sigma^{\dagger \dagger}$ &
  ${\lambda_1}^{*}$ & ${\lambda_2}^{*}$ & 
$\chi^{2}_{\nu}$ \\
No. &           & No.            &            &                            &      &
                            &                      &                           &
  \AA               &     \AA           &
            \\
\tableline
% \vspace{1ex}
%\vspace{5pt}
1   & Galactic         & 1    & 0.227 & 0.003 & 2.51 & 0.03 & $-$1.75 & 0.01 & 2500 & 8700 & 2.67 \\
2   & LMC              & 2    & 0.281 & 0.004 & 3.16 &  -   & $-$1.40 & 0.02 & 2500 & 8700 & 5.31 \\
3   & SMC              & 2    & 0.161 & 0.006 & 2.93 &  -   & $-$1.63 & 0.03 & 2500 & 8700 & 14.2 \\
4   & LMC              & 2    & 0.396 & 0.005 & 3.16 &  -   & $-$0.72 & 0.03 & 2800 & 6000 & 2.50 \\
5   & mod. Gal.$^{**}$ & 3    & 0.211 & 0.002 & 3.08 &  -   & $-$1.76 & 0.01 & 1690 & 8700 & 3.08 \\
6   & Gal.+LMC$^{\ddagger \ddagger}$
                       & 1+2  & 0.267 & 0.003 & 2.95 & 0.02 & $-$1.60 & 0.01 & 1690 & 8700 & 3.72 \\
7   & Galactic         & 1    & 0.223 & 0.003 & 2.43 & 0.03 & $-$1.75 & 0.01 & 2549 & 8700 & 2.44 \\
7b  &                  &      &       &       &      &      & $+$0.28 & 0.06 & 1690 & 2549 &      \\
\tableline
\end{tabular}
\end{center}
\noindent{$\dagger$ Reference for extinction formulae: 1. CCM; 2. Pei (1982); 3 Pei (1982)
modified to allow a variable far--UV extinction component.  $R_V$ is a fixed value in the models
using the Pei (1982) formulae for extinction and a variable in the models using the CCM formulae.}

\noindent{$\dagger \dagger$ Only the formal error -- systematic errors probably dominate and
are much larger.}

\noindent{$\ddagger$ Power law spectral index for the continuum modeled as $f_{\nu} \propto \nu^{\alpha}$.
For model 7 a broken power law was used and the
continuum parameters for the blue part are given in line 7b.}

\noindent{$*$ Beginning and end wavelength of the spectral region modeled as a power law and
included in the fit.}

\noindent{$**$ A modified Galactic extinction with a reduced far--UV component (see text).}

\noindent{$\ddagger \ddagger$ A model with both Galactic and LMC type absorbers at $z_a = 0.524$.
The best fit found required no LMC type absorption (E(B - V) $\le$ 0.005).}
% \hline
\end{table}
\end{document}